\documentclass[aps,prb,twocolumn,amsmath,amssymb,showpacs,superscriptaddress,notitlepage,longbibliography,floatfix]{revtex4-2}
\usepackage{amssymb}
\usepackage{amsmath}
\usepackage{amsfonts}
\usepackage{appendix}
\usepackage{subfigure}
\usepackage{dcolumn}
\usepackage{mathrsfs}
\usepackage{bm}
\usepackage{graphicx}
\usepackage{epsfig}
\usepackage{epstopdf}
\usepackage{color}
\usepackage{chngcntr}
\usepackage{balance}
\usepackage[dvipsnames]{xcolor}
\usepackage{calc}
\usepackage{natbib}
\usepackage[colorlinks=true,breaklinks=true,linkcolor=blue,anchorcolor=blue,citecolor=blue,urlcolor=blue]{hyperref}
\usepackage{lipsum}
\usepackage{stmaryrd}
\usepackage{sidecap}
\usepackage{booktabs}
\usepackage{multirow}

\begin{document}
	
\title{Spontaneous spin superconductor state in ABCA-stacked tetralayer graphene}
\author{Shuai Li}
\email{lishuai@suda.edu.cn}
\affiliation{School of Physical Science and Technology, Soochow University, Suzhou 215006, China.}
\affiliation{Institute for Advanced Study, Soochow University, Suzhou 215006, China.}
\author{Yuan-Hang Ren}
\affiliation{School of Physical Science and Technology, Soochow University, Suzhou 215006, China.}
\author{Ao-Long Li}
\affiliation{State Key Laboratory of Surface Physics and Department of Physics, Fudan University, Shanghai 200433, China}
\affiliation{Interdisciplinary Center for Theoretical Physics and Information Sciences (ICTPIS), Fudan University, Shanghai 200433, China.}
\author{Hua Jiang}
\email{jianghuaphy@fudan.edu.cn}
\affiliation{Interdisciplinary Center for Theoretical Physics and Information Sciences (ICTPIS), Fudan University, Shanghai 200433, China.}

\begin{abstract}
  We theoretically demonstrate a spontaneous spin superconductor (SC) state in ABCA-stacked tetralayer graphene, under sequential effects of electron-electron (e-e) and electron-hole (e-h) interactions. First of all, we examine the ferromagnetic (FM) exchange instability and phase diagram of the system induced by the long-range e-e interaction. At non- or low-doping levels, the interaction trends to stabilize a FM phase with the coexisting electron and hole carriers. Superior to bilayer and trilayer systems, tetralayer graphene has a larger FM phase region and spin splitting, making it more advantageous to realize the spin SC state. Subsequently, we prove that the FM phase becomes unstable when attractive e-h interaction is considered. As a consequence, the spin SC state can be spontaneously formed at low temperature, where spin-triplet exciton pairs act as the equivalent of Cooper pairs. We further develop a consistent BCS-type theory for the spin SC state in ABCA-stacked graphene. The predicted spin superconducting gap can reach about $7.0$ meV, with a critical temperature of about 45 K for non-doping system. At last, we demonstrated a spin-current Josephson effect in the ABCA-stacked graphene spin SC heterojunction. Our findings enrich the prospective spin SC candidate materials, illuminating more possibilities for achieving non-dissipative super-spintronics.
\end{abstract}

\maketitle

\section{Introduction}
Ever since its discovery\cite{DiscSC2010}, the development of superconductivity has witnessed remarkable advancements\cite{DiscSC2010,BCS1957,Joseph1962,SCcupr2000,TITopoSC2011,TopoSC2017,HTcSC2021,NobelSC2023,TeslaHTSC2019,ProappHTSC2021}. Theoretically, the physics of superconductors (SCs) are well understood by the fundamental BCS theory: Electrons combine into spin-singlet Cooper pairs, then condense into a charge superfluid at low temperature supporting persistent charge supercurrents\cite{BCS1957}. On the other hand, searching for spin superfluid possessing spin-polarized supercurrents is a key aspect in spintronics\cite{SpinSupcurr2011,Spincurrspintr2015,SCspintr2015}. The combination of superconductivity and spintronics opens a new research direction: super-spintronics, and offers a promising avenue for the application of new types of spintronic devices with a substantial decrease in power consumption\cite{Nobelspintr2008,TwoDspintr2019}.

The unusual ferromagnetic (FM) SC is believed to hold a spin superfluid with equal-spin triplet paring\cite{ProxSCferr2005,OddtrioSCferr2005,CoSCFerrURhGe2001,SCUCoGe2007,NearFerrSC2019}. However, this intrinsic triplet pairing state is rare and only identified in few materials\cite{CoSCFerrURhGe2001,SCUCoGe2007,NearFerrSC2019,SCSrRuOspintr2003,SrRuOpwave2009,Spintrip2021}. Fortunately, the spin-triplet excitonic condensate, dubbed as the spin SC, constitutes another prospective candidate for the realization of spin superfluid and non-dissipative spin transport \cite{spinSCFMgra2011,spinsctri2012,SpinSCv0grap2013,EIhydgrap2020,ElecMeissSpin2013,GLtypeSC2013,FlattripEI2021,EIflat2023,ExpSCNonspin2018,EImagper2017,SigSpintrip2023,MonoTMDTEI2020,SpintripTopo2024}. Although the spin SC is a charge insulator, a spin supercurrent can flow without dissipation, \textit{i.e.} zero spin resistance. In addition, the London-type equation\cite{spinSCFMgra2011,ElecMeissSpin2013} and Ginzburg-Landau-type theory\cite{GLtypeSC2013} for spin SCs are well established, unveiling intriguing electric "Meissner effect" and the spin-current Josephson effect. Interplayed with topology, the triplet spin SC might also be exploited as a topological SC bearing potentials for topological quantum computing without using Majorana fermions\cite{NonAbeSC2022,Renpro2023}.

Several promising materials are predicted to host spin-triplet excitonic condensate, including monolayer and trilayer graphene\cite{spinSCFMgra2011, spinsctri2012, SpinSCv0grap2013, EIhydgrap2020}, Kagome lattice systems\cite{FlattripEI2021, EIflat2023}, perovskites\cite{EImagper2017, SigSpintrip2023}, and topological insulators\cite{MonoTMDTEI2020, SpintripTopo2024}. In these systems, spin splitting and the presence of dual carriers can be induced externally by magnetic fields and ferromagnets, or intrinsically through Coulomb interactions. Subsequently, the spin-triplet exciton pairs could form and condense into a spin superfluid. 
However, the experimental confirmation of the spin SC states confronts challenges stemming from inherent limitations, such as short exciton lifetimes resulted from electron-hole (e-h) recombination\cite{Enradirecom1991,ExcRadilife2015}, inadequate spin split by weak magnetic exchange interaction\cite{Spinferrgrap2008}, and difficulties in material synthesis \cite{FlattripEI2021, EIflat2023}. These constraints complicate the experimental validation of the spin SCs.
Encouragingly, the recently reported ABCA-stacked tetralayer (ABCA) graphene \cite{MoireABCA2021,Spontetragrap2024,CIABCA2024} might serve as a promising platform to overcome these drawbacks. Compared to bilayer and trilayer graphene, its notable flatter bands and more quenched kinetic energy can possibly enhance the effects of Coulomb interaction\cite{Tunintrgap2018}, leading to a larger intrinsic spin split. Additionally, the generated hole carriers reside above the electronic states, thereby inhibiting e-h recombination and endowing excitons with long lifetimes\cite{spinSCFMgra2011}. Therefore, the benefits of ABCA graphene could possibly make it an exceptional platform for achieving spin superconductivity. Given the experimental accessibility of high-quality samples, this system merits further theoretical and experimental explorations of the triplet spin SC state.

In this paper, we theoretically demonstrate that the ABCA graphene can realize a spontaneous spin SC state by sequentially considering the intrinsic interactions. At first, we study the FM exchange instability induced by long-range electron-electron (e-e) interaction using a variational method. In comparison with the bilayer and trilayer graphene, the phase diagram of tetralayer system shows an obviously lager FM phase region as a result of its divergent density of states (DOS) near the Dirac points. For non-doping or weakly doped systems, the interaction trends to stabilize a FM phase with the coexistence of electron and hole carriers.

Subsequently, we verify that the FM phase will not be stable with the consideration of attractive e-h interaction. Electrons and holes spontaneously bind into spin-triplet exciton pairs and then condense into a spin SC state. A consistent BCS-type theory of the spin SC state is developed. For non-doping system, the estimated spin superconducting gap and critical temperature can reach about $7.0$ meV and $45$ K, respectively. Given the successful fabrication of ABCA graphene in recent experiments, we expect this spin SC state to be identified experimentally by finely tuning the system parameters.
Finally, we demonstrate the behavior of spin-current Josephson effect in the ABCA graphene spin SC heterojunction.

The remainder of this paper is organized as: In Sec. \ref{model}, we describe the model Hamiltonian and the variational wave function method to study exchange instability. In Sec. \ref{phase}, we study the FM instability of the system and derive its phase diagram. Sec. \ref{BCS} is devoted to the verification of instability of the FM phase in the presence of e-h interaction. A BCS-type theory is further established for the spin SC state in ABCA graphene. In Sec. \ref{spinJoe}, we study the related spin-current Josephson effect. A brief conclusion of our work is given in Sec. \ref{conclu}. All supplementary materials are relegated into the appendices.

\section{Model and variational method}\label{model}
\begin{figure}
	\centering
	\includegraphics[width=0.95\columnwidth]{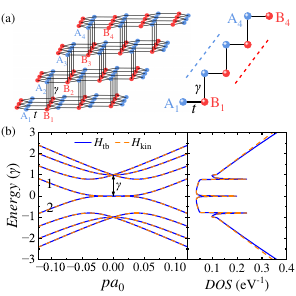}
	\caption{(Color online) (a) Schematic lattice structure of ABCA-stacked tetralayer graphene. (b) Numerical energy bands (left) along the $p_x$ axis near a valley $\boldsymbol{K}_{\eta}$ (set as $p=0$) and the corresponding DOS (right) calculated by diagonalizing $H_{\mathrm{tb}}$ and $H_{\mathrm{kin}}$, respectively.}
	\label{fig1}
\end{figure}

The lattice structure of ABCA graphene in one primitive cell is illustrated in Fig. \ref{fig1}(a). We model the kinetic energy of the system by a tight-binding Hamiltonian
\begin{equation}
	H_{\mathrm{tb}}= -t\sum_{\langle ij\rangle l\sigma} c^{\dagger}_{A_l i\sigma}c_{B_l j\sigma}  
	        -\gamma \sum_{i l\sigma} c^{\dagger}_{B_l i\sigma}  c_{A_{l+1} i\sigma}  + \mathrm{H.c.},
\end{equation}
where the operator $c_{\alpha_l i\sigma}$ annihilates an electron with spin $\sigma(\uparrow,\downarrow)$ at the site $i$ of sublattice $\alpha(\mathrm{A,B})$ on layer $l (1,\cdots,4$), $t\approx 3$ eV and $\gamma\approx 0.35$ eV are the nearest neighbor (NN) in-plane and out-of-plane hoppings, respectively\cite{eleelebil2006,spinsctri2012}. Perform an expansion of $H_{\mathrm{tb}}$  around $\boldsymbol{K}_{+}$ or $\boldsymbol{K}_{-}$ point at the corners of the Brillouin zone (BZ), then the low-energy kinetic Hamiltonian can be obtained as
\begin{equation}\label{Hkin}
	H_{\rm kin}=\sum_{\boldsymbol{p}}\sum_{\eta=+,-}\sum_{\sigma=\uparrow,\downarrow} \Psi^{\dagger}_{\eta\sigma}(\boldsymbol{p}) K_{\eta}(\boldsymbol{p})\Psi_{\eta\sigma}(\boldsymbol{p}).
\end{equation}
Here $\Psi_{\eta\sigma}(\boldsymbol{p})=[\psi_{A_1 \eta\sigma}(\boldsymbol{p}),\psi_{B_1 \eta\sigma}(\boldsymbol{p}),\cdots,\psi_{B_4 \eta\sigma}(\boldsymbol{p})]^{T}$, where $\psi_{\alpha \eta\sigma}(\boldsymbol{p})$ is the operator that annihilates an electron with spin $\sigma (\uparrow,\downarrow)$ on sublattice $\alpha(A_1,\cdots,B_4)$ of a small momentum $\boldsymbol{p}$ measured from valley $\boldsymbol{K}_\eta$ ($\eta=\pm$).
\begin{equation}\label{Kap}
	K_{\eta}(\boldsymbol{p})=
	\left(
	\begin{array}{cccc}
		H^{0}_{\eta}(\boldsymbol{p}) & t_{\bot} & 0 &0\\
		t_{\bot}^{\dagger}& H^{0}_{\eta}(\boldsymbol{p})&t_{\bot} &0\\
		0&t_{\bot}^{\dagger}& H^{0}_{\eta}(\boldsymbol{p})&t_{\bot}\\
		0& 0 & t_{\bot}^{\dagger} & H^{0}_{\eta}(\boldsymbol{p})\\
	\end{array}
	\right)
\end{equation}
gives the valley dependent Hamiltonian in momentum space, where $t_{\bot}=(0~0;-\gamma~0)$
is the interlayer coupling matrix and  $H^{0}_{\eta}(\boldsymbol{p})$
denotes the Hamiltonian of single layer graphene at long wavelength limit, truncated to the first order in $\boldsymbol{p}$ \cite{eleprograp2009}
\begin{equation}
	H^{0}_{\eta}(\boldsymbol{p})=\left(
	\begin{array}{cc}
		0 & \hbar\upsilon_{F}pe^{i \eta\theta(\boldsymbol{p})}\\
		\hbar\upsilon_{F}pe^{-i \eta\theta(\boldsymbol{p})}&0\\
	\end{array}
	\right),
\end{equation}
with $\tan\theta(\boldsymbol{p})=\frac{p_y}{p_x}$ and $\hbar\upsilon_F=\frac{3ta_0}{2}$ ($a_0\approx 0.142$ nm is the nearest carbon-carbon distance).

By solving the Schr\"odinger equation $K_{\eta}(\boldsymbol{p}) M_{\eta}(\boldsymbol{p})=M_{\eta}(\boldsymbol{p}) \varepsilon(\boldsymbol{p})$, the kinetic Hamiltonian is  diagonalized as
\begin{equation}
	H_{\rm kin}= \sum_{\boldsymbol{p} \eta\sigma} \Phi^{\dagger}_{\eta\sigma}(\boldsymbol{p}) \varepsilon(\boldsymbol{p}) \Phi_{\eta\sigma}(\boldsymbol{p})
	= \sum_{\boldsymbol{p}m \eta\sigma} \varepsilon_{m}(\boldsymbol{p}) \phi^{\dagger}_{\boldsymbol{p}m \eta\sigma} \phi_{\boldsymbol{p}m \eta\sigma},
\end{equation}
where $\Phi_{\eta\sigma}(\boldsymbol{p})= M^{\dagger}_{\eta}(\boldsymbol{p}) \Psi_{\eta\sigma}(\boldsymbol{p})=[\cdots,\phi_{\boldsymbol{p}m \eta\sigma},\cdots]^{T}$ with $m$ denoting the band index. We numerically plot the energy bands of ABCA graphene near $\boldsymbol{K}_{\eta}$ in Fig. \ref{fig1}(b), where the lowest-energy conduction and valence bands (labeled 1 and 2) intersect at the Dirac point ($E=0$) with quartic dispersion $\varepsilon_{1,2}(\boldsymbol{p})\approx\pm\frac{(\hbar\upsilon_Fp)^4}{\gamma^3}$ \cite{Fomulgrap2016}.

The low-energy Hamiltonian $H_{\mathrm{kin}}$ is only valid when the momentum $p$ is smaller than some critical value $k_c$ \cite{eletrans2011,spinferrtri2011}. Thus, we have to guarantee that the physics of ABCA graphene should mostly depend on the low-energy region and treat high-energy bands as non-important approximations. The reason why $H_{\mathrm{kin}}$ is used instead of $H_{\mathrm{tb}}$ is that it endows a rigid rotation symmetry around valleys $\boldsymbol{K}_{\eta}$, thereby making numerical calculations easier to implement.
To explain that $H_{\mathrm{kin}}$ can depict the low-energy physics as well as $H_{\mathrm{tb}}$, we numerically calculate their energy bands and DOS, as shown in Fig. \ref{fig1}(b). In the region of $|E|\leq0.5\gamma$ and $pa_0\leq0.1$, the bands (labeled 1 and 2) and DOS given by $H_{\mathrm{kin}}$ coincide well with that given by $H_{\mathrm{tb}}$, and are accurate in tolerance within 2\%. Since the Fermi energy and momentum invoked in this work are well located in this region, $H_{\mathrm{kin}}$ is then applicable in calculations.

Below we study the kinetic energy density of a given variational state, which can be calculated by
\begin{equation}\label{kinden}
	\frac{E_{\rm kin}}{S}= \frac{\langle H_{\rm{kin}} \rangle}{S}=\int_{D} \frac{d^2\boldsymbol{p}}{(2\pi)^2} ~ \sum_{m\eta\sigma}
	\varepsilon_{m}(\boldsymbol{p}) n_{m\eta\sigma}(\boldsymbol{p}),
\end{equation}
where $S$ represents the area of the system, $n_{m\eta\sigma}(\boldsymbol{p})=\langle \phi^{\dagger}_{\boldsymbol{p}m\eta\sigma}\phi_{\boldsymbol{p}m\eta\sigma} \rangle=0$, 1, $\Theta(Q_\sigma-p)$, or $1-\Theta(Q_\sigma-p)$ when the band $m$ with spin $\sigma$ at valley $\boldsymbol{K}_{\eta}$ is empty, fully occupied, partially filled by electrons, or by holes to momentum $Q_\sigma$ (pocket size). Here $Q_\sigma$ is the variational parameter, and
$\Theta$ is the Heaviside step function. The integral is performed over a disk area $D$ centered at $\boldsymbol{K}_{\eta}$ with radius $\Lambda=\sqrt{\frac{4\pi}{3\sqrt{3}a^2_0}}$, which ensures that the number of states is conserved in the first BZ \cite{eleelebil2006}.

Next we consider the exchange interaction between electrons. Because only weakly doped system is studied, the Coulomb interaction should be slightly screened and thus long ranged \cite{eeIntgrap2012,Polscrmulgrap2012}. Therefore, the e-e interaction can be given by $V(\boldsymbol{r}-\boldsymbol{r}')=\frac{e^2}{\epsilon|\boldsymbol{r}-\boldsymbol{r}'|}$, where $\epsilon$ is the dielectric constant of the substrate ($\epsilon\approx 1$ for suspended samples).
In the framework of second quantization, the e-e exchange interaction can be written in terms of field operators $\hat{\varphi}(\boldsymbol{r})$ as
\begin{equation}\label{secHex}
	H_{\rm ex}=\frac{1}{2}\int \int d\boldsymbol{r} d\boldsymbol{r}' ~\hat{\varphi}^{\dagger}(\boldsymbol{r})\hat{\varphi}^{\dagger}(\boldsymbol{r}') 
	V(\boldsymbol{r}-\boldsymbol{r}')  \hat{\varphi}(\boldsymbol{r}')\hat{\varphi}(\boldsymbol{r}).
\end{equation}
Expanding $\hat{\varphi}(\boldsymbol{r})$ in the basis of orbital states depicted by $\psi_{\alpha a\sigma}(\boldsymbol{p})$, 
$H_{\rm ex}$ can be recast to (see Appendix \ref{appdex} for derivation) \cite{eleelebil2006,spinsctri2012}
\begin{equation}\label{chardensHex}
	H_{\rm ex}= \frac{1}{2S}\sum_{\boldsymbol{q}\neq\boldsymbol{0}}\sum^{4}_{i,j=1} \hat{\rho}_{i}(\boldsymbol{q})  V_{ij}(\boldsymbol{q}) \hat{\rho}_{j}(-\boldsymbol{q}),
\end{equation}
where $V_{ij}(\boldsymbol{q})=2\pi e^2 e^{-qd_{ij}}/\epsilon q= 2\pi g\hbar \upsilon_{F} e^{-qd_{ij}}/q$ gives the interaction between layers $i$ and $j$, with $d_{ij}$ the vertical distance between the two layers and $g=e^2/(\epsilon \hbar\upsilon_{F})$ defined as the dimensionless interaction governing the system and $\hat{\rho}_{i}(\boldsymbol{q})=\sum_{\alpha\boldsymbol{p} a\sigma}  
\psi^{\dagger}_{\alpha_i a\sigma}(\boldsymbol{p}+\boldsymbol{q}) \psi_{\alpha_i a\sigma}(\boldsymbol{p})$ is the Fourier component of layer charge density $\hat{\rho}_{i}(\boldsymbol{r}_i)$.

The presence of long-range interaction described by Eq. \eqref{chardensHex} can lead to the exchange instability in the noninteracting ground state of ABCA graphene. We will utilize a variational wave function technique to study the FM instability, which is important in 2D electron gas with diluted carriers\cite{coulferrgra2005,eleelebil2006,spinsctri2012,spinferrtri2011,Ferromag1947}. First, it is convenient to convert $\hat{\rho}_{i}(\boldsymbol{q})$ into the diagonalized basis by
\begin{equation}\label{diagdensity}
	\hat{\rho}_{i}(\boldsymbol{q})=\sum_{\boldsymbol{p} \eta\sigma} \Phi^{\dagger}_{\eta\sigma}(\boldsymbol{p}+\boldsymbol{q})~ \chi^{i}(\boldsymbol{p}+\boldsymbol{q},\boldsymbol{p})~  \Phi_{\eta\sigma}(\boldsymbol{p})
\end{equation}
where
\begin{equation}
  \chi^{i}(\boldsymbol{p}+\boldsymbol{q},\boldsymbol{p})=M_{\eta}(\boldsymbol{p}+\boldsymbol{q})^{\dagger} \mathrm{diag}(\cdots,0,1,1,0,\cdots)~ 
  M_{\eta}(\boldsymbol{p}).
\end{equation}
Then substituting $\hat{\rho}_{i}(\boldsymbol{q})$ in Eq. \eqref{chardensHex} with Eq. \eqref{diagdensity} yields the exchange energy density for a chosen variational state
\begin{equation}\label{exden}
\begin{split}
     &\frac{E_{\mathrm{ex}}}{S} = \frac{\langle H_{\mathrm{ex}}\rangle}{S} \approx -\frac{1}{2} \int_D \frac{d^2\boldsymbol{p}}{(2\pi)^2} \int_D \frac{d^2\boldsymbol{q}}{(2\pi)^2} \sum_{i,j} \sum_{m,n} \sum_{\eta,\sigma} \\
	  & \chi^{i}_{mn,\eta}(\boldsymbol{q},\boldsymbol{p}) \chi^{j}_{nm,\eta}(\boldsymbol{p},\boldsymbol{q}) V_{ij}(\boldsymbol{p}-\boldsymbol{q}) 	n_{m\eta\sigma}(\boldsymbol{q}) n_{n\eta\sigma}(\boldsymbol{p}),
\end{split}
\end{equation}
where the scattering between states of the two different valleys has been dropped since their momenta difference $\boldsymbol{p}-\boldsymbol{q}$ is very large to render $V_{ij}(\boldsymbol{p}-\boldsymbol{q})$ to be extremely small and thus safely neglected. Note that $E_{\mathrm{kin}}/S$ and $E_{\mathrm{ex}}/S$ are all functions of the variational parameter $Q_{\sigma}$, as shown by Eqs. \eqref{kinden} and \eqref{exden}.

The total energy density for the ABCA graphene is given by $E/S=(E_{\mathrm{kin}}+E_{\mathrm{ex}})/S$. To determine the pocket size of the equilibrium state, we have to minimize $E/S$ with respect to $Q_\sigma$ to find $Q_{\sigma,\mathrm{min}}$, for which the energy density is minimized.


\begin{SCfigure*}
	\centering
	\includegraphics[width=0.65\textwidth]{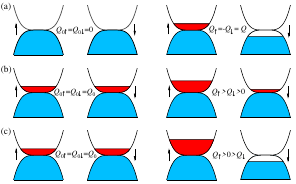}
	\quad \qquad
	\caption{(Color online) (a) Half-filling case. Left: noninteracting ground state with zero pockets. Right: variational state with one electron pocket $Q_{\uparrow}$ in spin up and one hole pocket $Q_{\downarrow}$ in spin down channels. (b-c) Low-doping cases. Left: noninteracting ground state with two electron pockets $Q_{0\uparrow}=Q_{0\downarrow}$. Right in (b): variational state with two asymmetric electron pockets $Q_{\uparrow}>Q_{\downarrow}>0$; Right in (c): variational state with one electron pocket $Q_{\uparrow}$ and one hole pocket $Q_{\downarrow}$. At the two valleys $\boldsymbol{K}_{+}$ and $\boldsymbol{K}_{-}$, the pockets are degenerate.
    \vspace{9mm}}
	\label{fig2}
\end{SCfigure*}

\section{FM exchange instability and phase diagram}\label{phase}
As mentioned above, the FM exchange instability plays important roles in electronic systems with low density of carriers\cite{Ferromag1947}. In this section, we will study the FM instability in weakly doped ABCA graphene with long-range Coulomb interaction and acquire its phase diagram at zero temperature, utilizing the variational method.

\begin{figure}
	\centering
	\includegraphics[width=1.0\columnwidth]{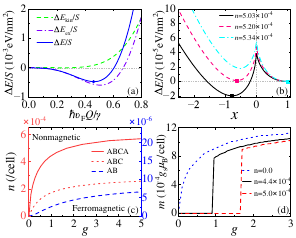}
	\caption{(Color online) (a) The variation of kinetic energy density $\Delta E_{\mathrm{kin}}/S$, exchange energy density $\Delta E_{\mathrm{ex}}/S$ and total energy density $\Delta E/S$ vs $\hbar\upsilon_{F}Q/\gamma$ for non-doping system ($n=0$) with $g=2.1$. Blue dot labels the minimum of $\Delta E/S$. (b) $\Delta E/S$ vs $x=s_{\downarrow}Q^4_{\downarrow}/Q^4_{0}$ for low-doping systems with different $n$ and $g=2.1$. Squares label minima of $\Delta E/S$. (c) Phase diagrams vs $n$ and $g$ for ABCA, ABC and AB graphene systems. The three lines are boundaries separating FM and NM phases. (d) Equilibrium FM magnetization $m$ vs $g$ for half filling ($n=0$) and doped ($n\neq 0$) ABCA graphene.}
	\label{fig3}
\end{figure}

First, let us start with a non-doping system with $n=0$ (here $n$ is the number of carriers in one primitive cell doped away from the Dirac point). In the noninteracting ground state,  the bands of the two spin channels are filled to the Dirac point with zero pockets $Q_{0\uparrow}=Q_{0\downarrow}=0$. Under e-e interaction, the FM variational state can be chosen with one electron pocket in the spin-up branch ($Q_{\uparrow}>0$) and one hole pocket in the spin-down branch ($Q_{\downarrow}<0$) at each valley, such that the sizes of the pockets satisfy $Q_{\uparrow}=-Q_{\downarrow}=Q$, because of the conservation of particle number in the non-doping system. The noninteracting ground state and the variational trial state are schematically shown in Fig. \ref{fig2}(a). To gain the properties of equilibrium state, we calculate the energy density variations $\Delta E_{\mathrm{kin}}/S$ and $\Delta E_{\mathrm{ex}}/S$ relative to noninteracting ground state as functions of $Q$, which are potted in Fig. \ref{fig3}(a). 
Analytically, the kinetic energy cost by an electron (hole) pocket is given by the integral $\int_{0}^{E(Q)} \rho(E) E dE$. Since the DOS for the quartic band is $\rho(E) \sim 1/\sqrt{E}$ and $E\sim Q^4$, hence $\Delta E_{\mathrm{kin}}/S\sim Q^6$, as consistent with the numerical results in Fig. \ref{fig3}(a). On the other hand, $\Delta E_{\mathrm{ex}}/S$ exhibits a nonmonotonic behavior with $Q$, which first decreases to the minimum and then goes up. 
Therefore, the sum of $\Delta E_{\mathrm{kin}}/S$ and $\Delta E_{\mathrm{ex}}/S$ gives rise to $\Delta E/S$ showing a minimum at some specific $Q_{\mathrm{min}}$. For the unscreened system ($\epsilon=1$ and $g=2.1$ inherent of the ABCA graphene), $\Delta E/S$ reaches the minimum [labeled by the blue dot in Fig. \ref{fig3}(a)] with a spin-split energy $\hbar\upsilon_{F}Q_{\mathrm{min}}\approx 0.46 \gamma$ of the equilibrium state. Therefore, the e-e interaction stabilizes a spin-polarized FM phase in the non-doping ABCA graphene where electron and hole pockets form simultaneously, with a magnetization $m=n_{\uparrow}-n_{\downarrow}\approx 1.10\times 10^{-3} g_s\mu_B$ in one cell. The spin-split energy and magnetization significantly exceed that reported in bilayer  ($\hbar\upsilon_{F}Q_{\mathrm{min}}\approx 0.05\gamma$ \cite{eleelebil2006}) and trilayer ($\hbar\upsilon_{F}Q_{\mathrm{min}}\approx 0.09\gamma$ \cite{spinsctri2012}) graphene, which arises from the flatter energy bands ($\sim k^4$) and divergent DOS of the tetralyer graphene system near the Dirac point.

In realistic experiment, the interaction $g$ can be flexibly tuned by external substrates\cite{UntordMAgrap2020,Tunelecor2021}. Here, we explore the effect of $g$ on the equilibrium states. As shown in Fig. \ref{sfig1} in Appendix \ref{appendhf}, $\Delta E_{\mathrm{ex}}/S$ is lowered while $\Delta E_{\mathrm{kin}}/S$ keeps invariant as $g$ increases. The pocket size $Q_{\mathrm{min}}$ of the equilibrium state gradually goes up and saturates for large enough $g$. Since FM magnetization $m\propto Q^2_{\mathrm{min}}$, $m$ displays a similar behavior with $g$ as shown by the blue line in Fig. \ref{fig3}(d). For $g=3.0$, $m$ can reach a notable value $1.10\times 10^{-3} g_s\mu_B$, about two orders larger than that in bilayer graphene \cite{eleelebil2006}. It is now clear from above study that a FM phase is always energetically favorable for any small $g$ in the non-doping ABCA graphene system. This is also confirmed and illustrated by the $n=0$ line in the phase diagram shown in Fig. \ref{fig3}(c).

The magnetization of equilibrium states should not have a spin preference, since the two spin channels are degenerate and scattering between them are forbidden [see Eq. \eqref{exden}]. We present a quantitative study in Appendix \ref{appendhf} to elucidate that the spin-flipped state is degenerate with the FM state in Fig. \ref{fig2}(a) at equilibrium.
Note that the selected FM variational state above preserves $\mathrm{Z}_2$ symmetry but with broken $\mathrm{SU}(2)$ symmetry, hence it is spin polarized\cite{eleelebil2006}. There is another possible state which breaks both symmetries but without net magnetization: The spin up (down) channels at the two valleys have reversed pockets and the two spin channels are opposite as well at the same valley [see Fig. \ref{sfig2}(c)]. Because the intervalley scattering is neglected, this nonmagnetic (NM) state is also degenerate with the FM state in Fig. \ref{fig2}(a). Thus, it is necessary to incorporate the intervalley scattering contribution to determine which state is more energetically favorable. A detailed discussion is given in Appendix \ref{appendhf}, where we find that the FM phase has a smaller energy density than the NM phase, and thus is more stable.

Next, we investigate the FM instability in weakly doped ABCA graphene. Suppose that the noninteracting ground state is initially doped with electron pockets $Q_{0\uparrow}=Q_{0\downarrow}=Q_0$ for the two spin channels, as shown in the left panels of Figs. \ref{fig2}(b-c). When considering e-e interaction, FM instability occurs. There are two types of FM variational states in the doped case, as displayed in the right panels of \ref{fig2}(b-c): One type with two asymmetric electron pockets ($Q_\uparrow>Q_\downarrow>0$), and the other type with one electron pocket and one hole pocket ($Q_\uparrow>0>Q_\downarrow$) for the spin up and spin down channels, respectively. With conserved number of particles, the pocket sizes are constrained by
\begin{equation}
	s_{\uparrow}\frac{Q^2_{\uparrow}}{4\pi}+s_{\downarrow}\frac{Q^2_{\downarrow}}{4\pi}=
	s_{0\uparrow}\frac{Q^2_{0\uparrow}}{4\pi}+s_{0\downarrow}\frac{Q^2_{0\downarrow}}{4\pi} = \frac{n}{S_{\mathrm{cell}}}
\end{equation}
where $s_{(0)\uparrow/\downarrow}$ equals +1 for electron pockets, and  -1 for hole pockets, and $S_{\mathrm{cell}}$ is the area of one primitive cell. From above, we see that there is only one independent variational parameter that accounts. Here we choose $Q_{\downarrow}$ without loss of generality. Note that there is always an electronlike pocket for the FM variational state [see Figs. \ref{fig2}(b-c)], therefore we generically set $s_{\uparrow}=1$. Also, the spin-resolved pockets are imbalanced with $Q_{\uparrow}\neq |Q_{\downarrow}|$.

Below we evaluate the energy density $\Delta E/S$ vs $x=s_{\downarrow}Q^4_{\downarrow}/Q^4_{0}$ to study the equilibrium states. The calculation results for $g=2.1$ with different doping constraints $n$ are plotted in Fig. \ref{fig3}(b). For slightly doping $n=5.03\times 10^{-4}$ (see black solid line), the equilibrium state is FM phased with a hole pocket in the spin down channel (denoted by the black square: $x<0\Rightarrow Q_{\downarrow}<0$) and also an electron pocket in spin up channel. When doping is enhanced to  $n=5.34\times 10^{-4}$ (see cyan dash-dot line), the system is transformed to NM phase at equilibrium (denoted by the cyan square: $x=1\Rightarrow Q_{\downarrow}=Q_{\uparrow}=Q_0$). There is a transition point where the FM phase is degenerate with the NM phase at equilibrium when $n=5.20\times 10^{-4}$ (see pink dash line and squares).
By now, we numerically find that there exist two types of exchange instabilities, the FM and NM instabilities, in doped ABCA graphene. To systematically study the equilibrium states, we have calculated the phase diagram as functions of $g$ and $n$, as given in Fig. \ref{fig3}(c). At small $n$, the system trends to develop a FM phase even with small $g$. For moderate $n$, the FM phases can be stabilized by increasing $g$. Whereas it can not stay magnetic anymore when $n$ exceeds some critical threshold, even $g$ is significantly large. The FM phases are all characterized with one electron pocket in one spin channel and one hole pocket in the other, rather than two electronlike pockets [see Figs. \ref{fig2}(b-c)].  We further note that the vanishing magnetization $m$ in the NM state dramatically changes to finite once $g$ crosses the transition line, as shown in Fig. \ref{fig3}(d). And $m$ will be suppressed as $n$ increases.

Moreover, we also studied the layer effect on the phase diagram. The results for AB bilayer and ABC trilayer graphene are displayed in Fig. \ref{fig3}(c) to give comparison with the ABCA graphene. We see that the FM phase region of the tetralayer system is much larger than that of the bilayer and trilayer systems. This is attributed to the peculiar quartic spectrum near the Dirac point, which gives rise to divergent DOS [see Fig. \ref{fig1}(b)]. FM phase can be more favorable and easily generated according to the Stoner criteria \cite{Ferromag1947}. Thus, the ABCA graphene is more flexible to exploit with the FM phases, superior to bilayer and trilayer graphene systems.

In summary, we have studied the FM exchange instability of the weakly doped ABCA graphene in the presence of long-range e-e interaction and acquired zero temperature phase diagram. At non- or low-doping, the interaction prefers to stabilize spin-polarized FM phases at equilibrium, where an electronlike pocket and a holelike pocket are formed at the same time for the two spin channels. The ABCA graphene will be more advantageous than the bilayer and trilayer systems in the flexibility on manipulating FM phases, due to a larger FM-phase region of its phase diagram.

\section{Spontaneous spin superconductor state}\label{BCS}
\subsection{Instability of the FM state}
The intrinsic e-e interaction will result in a FM state in ABCA graphene with the coexistence of electron and hole pockets. Below we will prove that the naturally occurring attractive interaction between electron and hole carriers, no matter how weak it is, will lead to the instability of this FM state. Electrons and holes will spontaneously bound into spin-triplet exciton pairs.

\begin{figure}
	\centering
	\includegraphics[width=1.0\columnwidth]{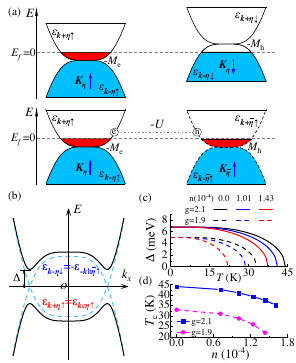}
	\caption{(Color online) (a) Top panel: Schematic of low-energy bands of FM phased ABCA gaphene for spin-up (left) and spin-down (right) electrons in valley $\boldsymbol{K}_\eta$. Bottom panel: Low-energy bands of FM ABCA graphene for spin-up electrons in valley $\boldsymbol{K}_\eta$ (left) and spin-up holes in opposite valley $\boldsymbol{K}_{\bar{\eta}}$ (right). An additional e-h pair with attractive interaction $-U$ is labeled. The horizontal dashed lines mark the Fermi levels $E_f=0$. (b) Schematic of low-energy bands of FM phased ABCA graphene for free fermions with pair potential $\Delta=0$ (dashed lines), and for excitonic condensate with $\Delta \neq 0$ (solid lines). (c) $\Delta$ vs $T$ with different $g$ and $n$. (d) $T_c$ vs $n$ with different $g$.}
	\label{fig4}
\end{figure}

At the low energies of the FM state of ABCA graphene, the noninteracting Hamiltonian for the spin $\uparrow$ (electron-like) and spin $\downarrow$ (hole-like) subsystems in valley $\boldsymbol{K}_\eta(\eta=\pm)$ can be approximately written as
\begin{equation}
	H^0_{\uparrow(\downarrow)}=\left(
	\begin{array}{cc}
		- M_{e(h)} & \frac{\hbar^4\upsilon^4_F(\hat{k}^{e}_{x}+i\hat{k}^{e}_{y})^4}{\gamma^3}\\
		\frac{\hbar^4\upsilon^4_F(\hat{k}^{e}_{x}-i\hat{k}^{e}_{y})^4}{\gamma^3}&- M_{e(h)} \\
	\end{array}
	\right),
\end{equation}
where $M_e>0$ ($M_h<0$) is the chemical potential of the spin $\uparrow$ ($\downarrow$) subsystem with respect to the Dirac point, and $\boldsymbol{\hat{k}}^{e}=(\hat{k}^{e}_{x}, \hat{k}^{e}_{y})$ is the momentum operator for the electronic states. We temporarily set $\hbar=1$ hereafter in this section.
Then, the corresponding Schr\"odinger equations for electrons of the two subsystems are given by
\begin{equation}\label{eleSchr}
	\begin{split}
		\left(
		\begin{array}{cc}
			-M_e & \frac{\upsilon^4_F(\hat{k}^{e}_{x}+i\hat{k}^{e}_{y})^4}{\gamma^3}\\
			\frac{\upsilon^4_F(\hat{k}^{e}_{x}-i\hat{k}^{e}_{y})^4}{\gamma^3}&-M_e\\
		\end{array}
		\right) \varphi^{0e}_{\boldsymbol{k}n\eta\uparrow}(\boldsymbol{r}_e)=& \varepsilon_{\boldsymbol{k}n\eta\uparrow} \varphi^{0e}_{\boldsymbol{k}n\eta\uparrow}(\boldsymbol{r}_e); \\
		\left(
		\begin{array}{cc}
			-M_h & \frac{\upsilon^4_F(\hat{k}^{e}_{x}+i\hat{k}^{e}_{y})^4}{\gamma^3}\\
			\frac{\upsilon^4_F(\hat{k}^{e}_{x}-i\hat{k}^{e}_{y})^4}{\gamma^3}&-M_h\\
		\end{array}
		\right) \varphi^{0e}_{\boldsymbol{k}n\eta\downarrow}(\boldsymbol{r}_e)=& \varepsilon_{\boldsymbol{k}n\eta\downarrow} \varphi^{0e}_{\boldsymbol{k}n\eta\downarrow}(\boldsymbol{r}_e).
	\end{split}
\end{equation}
The eigenenergies and eigenstates for the spin up subsystem are solved as
\begin{equation}
\begin{split}
	&\varepsilon_{\boldsymbol{k}\boldsymbol{\pm} \eta\uparrow}= \pm \frac{\upsilon^4_Fk^4}{\gamma^3}-M_e; \\
	&\varphi^{0e}_{\boldsymbol{k}\boldsymbol{\pm} \eta\uparrow}(\boldsymbol{r}_e)= \frac{e^{i\boldsymbol{k}\cdot \boldsymbol{r}_e}}{\sqrt{2}}
	\left(
	\begin{array}{cc}
		1\\
		\pm e^{-i4\theta_{\boldsymbol{k}}}\\
	\end{array}
	\right),
\end{split}
\end{equation}
where $\theta_{\boldsymbol{k}}=\arctan \frac{k_y}{k_x}$. For spin down system, $\varepsilon_{\boldsymbol{k}\boldsymbol{\pm} \eta\downarrow}=\boldsymbol{\pm} \frac{\upsilon^4_Fk^4}{\gamma^3} -M_h$ and $\varphi^{0e}_{\boldsymbol{k}\boldsymbol{\pm} \eta\downarrow}(\boldsymbol{r}_e)=\varphi^{0e}_{\boldsymbol{k}\pm \eta\uparrow}(\boldsymbol{r}_e)$. We show the low-energy bands of the FM state in the top panel of Fig. \ref{fig4}(a). The spin-up conduction band are partially filled with electron-like carriers, while the spin-down carriers are hole-like due to the hole pocket in the valence band. We further take the e-h transformation for the spin down subsystem. The annihilation of an electron with spin down in valley $\boldsymbol{K}_{\eta}$ is equivalent to the creation of a spin up hole with reversed momentum and energy in the opposite valley $\boldsymbol{K}_{\bar{\eta}}=\boldsymbol{K}_{-\eta}$. By transformation of Eq. \eqref{eleSchr}, the Schr\"odinger equation for free holes can be obtained as
\begin{small}
	\begin{equation}\label{holeShro}
		-\left(
		\begin{array}{cc}
			-M_h & \frac{\upsilon^4_F(-\hat{k}^{h}_{x}+i\hat{k}^{h}_{y})^4}{\gamma^3}\\
			\frac{\upsilon^4_F(-\hat{k}^{h}_{x}-i\hat{k}^{h}_{y})^4}{\gamma^3}&-M_h\\
		\end{array}
		\right) \varphi^{0h}_{\boldsymbol{k}n\bar{\eta}\uparrow}(\boldsymbol{r}_h)= \varepsilon_{\boldsymbol{k}n\bar{\eta}\uparrow} \varphi^{0h}_{\boldsymbol{k}n\bar{\eta}\uparrow}(\boldsymbol{r}_h).
	\end{equation}
\end{small}
Its solutions are given by
\begin{equation}
\begin{split}
	&\varepsilon_{\boldsymbol{k}\boldsymbol{\pm} \bar{\eta}\uparrow}=\pm \frac{\upsilon^4_Fk^4}{\gamma^3} +M_h, \\
	&\varphi^{0h}_{\boldsymbol{k}\boldsymbol{\pm} \bar{\eta}\uparrow}(\boldsymbol{r}_h)=\frac{e^{i\boldsymbol{k}\cdot \boldsymbol{r}_h}}{\sqrt{2}}
	\left(
	\begin{array}{cc}
		1\\
		\mp e^{i4\theta_{\boldsymbol{k}}}\\
	\end{array}
	\right).
\end{split}
\end{equation}
The bottom panel of Fig. \ref{fig4}(a) shows the energy bands for the holes and electrons in different valleys. In the FM state of ABCA graphene, the lowest conduction bands of electrons and holes are filled up to the chemical potentials $M_\mathrm{e}$ and $-M_\mathrm{h}$, respectively.

Next, we study the instability of the FM state by taking e-h interaction into account. Following the theory of Ref. \cite{spinSCFMgra2011},  we consider an additional e-h pair residing out of the FM state composed with free fermions [as illustrated in the bottom panel of Fig. \ref{fig4}(a)]. The only effect of other free electrons and holes is to prohibit the e-h pair from occupying states below Fermi energy $E_f=0$. We will subsequently show that with attractive interaction $-U(\boldsymbol{r}_e,\boldsymbol{r}_h)$, this e-h pair spontaneously forms mixed state with neutral charge and $2(\hbar/2)$ spin. The Schr\"odinger equation for the interacting e-h pair reads
\begin{widetext}
\begin{equation}\label{ehSchr}
	\left[ \left( \begin{array}{cc}
    	- M_e & \frac{\upsilon^4_F(\hat{k}^{e}_{x}+i\hat{k}^{e}_{y})^4}{\gamma^3}\\
		\frac{\upsilon^4_F(\hat{k}^{e}_{x}-i\hat{k}^{e}_{y})^4}{\gamma^3}&- M_e\\
	     \end{array} \right) 
	  -\left( \begin{array}{cc}
		-M_h & \frac{\upsilon^4_F(-\hat{k}^{h}_{x}+i\hat{k}^{h}_{y})^4}{\gamma^3}\\
		\frac{\upsilon^4_F(-\hat{k}^{h}_{x}-i\hat{k}^{h}_{y})^4}{\gamma^3}& -M_h\\
		\end{array} \right) - U(\boldsymbol{r}_e,\boldsymbol{r}_h)  \right]
		\psi_{\uparrow}(\boldsymbol{r}_e,\boldsymbol{r}_h)=E\psi_{\uparrow}(\boldsymbol{r}_e,\boldsymbol{r}_h),
\end{equation}
\end{widetext}
where $\psi_{\uparrow}(\boldsymbol{r}_e,\boldsymbol{r}_h)$ is a two-particle wave function and can be expanded by
\begin{equation}
	\psi_{\uparrow}(\boldsymbol{r}_e,\boldsymbol{r}_h)=\sum_{\boldsymbol{k}} g_{\boldsymbol{k}} \varphi^{0e}_{\boldsymbol{k}\boldsymbol{+} \eta\uparrow}(\boldsymbol{r}_e) \varphi^{0h}_{-\boldsymbol{k}\boldsymbol{+}\bar{\eta}\uparrow}(\boldsymbol{r}_h),
\end{equation}
with the coefficient $g_{\boldsymbol{k}}=0$ if $\frac{\upsilon^4_F|\boldsymbol{k}|^4}{\gamma^3}-max(M_e,-M_h)\leq 0$ due to the Pauli exclusion principle. Note that only free electron and hole states with opposite momenta are paired since they are energetically favorable, and the mixed $\psi_{\uparrow}(\boldsymbol{r}_e,\boldsymbol{r}_h)$ has a nonzero spin $2(\hbar/2)$. Put $\psi_{\uparrow}(\boldsymbol{r}_e,\boldsymbol{r}_h)$ into Eq. \eqref{ehSchr},  we obtain
\begin{equation}
\begin{split}
	 (& 2\frac{\upsilon^4_F|\boldsymbol{k}|^4}{\gamma^3}- M_e + M_h) g_{\boldsymbol{k}} - \sum_{\boldsymbol{k}'} g_{\boldsymbol{k}'} U_{\boldsymbol{k}\boldsymbol{k}'}=Eg_{\boldsymbol{k}} \\
	&\Rightarrow ~~ g_{\boldsymbol{k}}= \sum_{\boldsymbol{k}'} \frac{g_{\boldsymbol{k}'U_{\boldsymbol{k}\boldsymbol{k}'}}}{2\frac{\upsilon^4_F|\boldsymbol{k}|^4}{\gamma^3}-M_e +M_h-E}
\end{split}
\end{equation}
where $U_{\boldsymbol{k}\boldsymbol{k}'}=\int\int d\boldsymbol{r}_e d\boldsymbol{r}_h ~ U(\boldsymbol{r}_e,\boldsymbol{r}_h) e^{-i(\boldsymbol{k}-\boldsymbol{k}')(\boldsymbol{r}_e-\boldsymbol{r}_h)}$.
Because that $U_{\boldsymbol{k}\boldsymbol{k}'}$ decreases slowly with $|\boldsymbol{k}-\boldsymbol{k}'|$, it is reasonable to assume it as a constant $U$, independent of $\boldsymbol{k}$ and $\boldsymbol{k}'$. Then the above equation reduces to
\begin{equation}
	1=-U\int_{max(M_e,-M_h)}^{\varepsilon_D} d\varepsilon ~~ \frac{\rho(\varepsilon)}{E+M_e -M_h-2\varepsilon},
\end{equation}
where $\rho(\varepsilon)\sim 1/\sqrt{\varepsilon}$ is the low-energy DOS of the bare ABCA graphene without interaction.
With positive $U$ and $M_e ~(-M_h)>0$, a negative $E$ solution of this equation is always possible. Therefore, we can prove that the mixed e-h pair is more stable since it is lower in energy than a free e-h pair (energy $E=2\frac{\upsilon^4_F|\boldsymbol{k}|^4}{\gamma^3}-M_e +M_h>0$). Electron and hole will spontaneously bind into exciton pair with neutral charge and spin polarization $2(\hbar/2)$. In conclusion, the FM state in ABCA graphene will be unstable in the presence of attractive e-h interaction.

\subsection{BCS-type theory for the spin superconductor}

In FM phased ABCA graphene, we have shown that electrons and holes will spontaneously bind into more stable spin-polarized exciton pairs. At low temperature, they could condense into a spin-polarized superconducting state, namely, a spin SC. Here, we develop a BCS-type theory for the spin SC state in ABCA graphene.

Let us begin by considering an interacting electron system in FM phased ABCA graphene at low energies. Its Hamiltonian reads $H=H_0+H_{\mathrm{int}}$, where $H_0$ represents the noninteracting Hamiltonian for electrons from the lowest conduction band (labeled $\boldsymbol{+}$) of spin up channel and the highest valence band (labeled $\boldsymbol{-}$) of the spin down channel [see Fig. \ref{fig4}(a)], and $H_{\mathrm{int}}$ depicts the e-e interaction between the two band branches:
\begin{equation}
\begin{split}
	& H_0=\sum_{\eta,\boldsymbol{k}} \varepsilon_{\boldsymbol{k}\boldsymbol{+} \eta\uparrow} \phi^{\dagger}_{\boldsymbol{k}\boldsymbol{+} \eta\uparrow} \phi_{\boldsymbol{k}\boldsymbol{+} \eta\uparrow} +
	\varepsilon_{\boldsymbol{k}\boldsymbol{-} \eta\downarrow} \phi^{\dagger}_{\boldsymbol{k}\boldsymbol{-} \eta\downarrow}\phi_{\boldsymbol{k}\boldsymbol{-} \eta\downarrow}, \\
	& H_{\mathrm{int}}= \sum_{\eta,\boldsymbol{k},\boldsymbol{k}',\boldsymbol{q}}  U^{\eta}_{\boldsymbol{k}\boldsymbol{k'}\boldsymbol{q}}
	\phi^{\dagger}_{\boldsymbol{k}-\boldsymbol{q}\boldsymbol{+} \eta\uparrow}  \phi^{\dagger}_{\boldsymbol{k}'+\boldsymbol{q}\boldsymbol{-} \eta\downarrow}
	\phi_{\boldsymbol{k}'\boldsymbol{-} \eta\downarrow}  \phi_{\boldsymbol{k}\boldsymbol{+} \eta\uparrow} .
\end{split}
\end{equation}
Here, the interaction matrix element $U^{\eta}_{\boldsymbol{k}\boldsymbol{k'}\boldsymbol{q}}=\langle \boldsymbol{k}-\boldsymbol{q}\boldsymbol{+},\eta\uparrow| \langle \boldsymbol{k}'+\boldsymbol{q}\boldsymbol{-} ,\eta\downarrow| V(\boldsymbol{r}-\boldsymbol{r}') |\boldsymbol{k}'\boldsymbol{-} ,\eta\downarrow \rangle  |\boldsymbol{k}\boldsymbol{+} ,\eta\uparrow \rangle$. The two band branches $\varepsilon_{\boldsymbol{k}\boldsymbol{+} \eta\uparrow}=\varepsilon_{\boldsymbol{k}}-M_e$ and $\varepsilon_{\boldsymbol{k}\boldsymbol{-} \eta\downarrow}=-\varepsilon_{\boldsymbol{k}} -M_h$ can be numerically calculated by the diagonalization of Eq. \eqref{Kap}, with their chemical potentials $M_e$ and $M_h$ determined by the variational method from last section.
Because the spin-down carrier is being hole-like in FM phased system, the annihilation of an electron by $\phi_{\boldsymbol{k}- \eta\downarrow}$ also means the creation of a hole with opposite momentum and spin. Therefore we can define the operator for a hole state by $\alpha^{\dagger}_{-\boldsymbol{k}h \bar{\eta}\uparrow}=\phi_{\boldsymbol{k}\boldsymbol{-} \eta\downarrow}$. For an electronic state in spin-up band, we define $\alpha_{\boldsymbol{k}e \eta\uparrow}=\phi_{\boldsymbol{k}\boldsymbol{+} \eta\uparrow}$.
Then the noninteracting Hamiltonian $H_0$ can be transformed to
\begin{small}
	\begin{equation}\label{H0SC}
		H_0=\sum_{\eta,\boldsymbol{k}} \left(
		\begin{array}{cc}
			\alpha^{\dagger}_{\boldsymbol{k}e \eta\uparrow} & \alpha_{-\boldsymbol{k}h \bar{\eta}\uparrow}\\
		\end{array}
		\right) \left(
		\begin{array}{cc}
			\varepsilon_{\boldsymbol{k}e \eta\uparrow} & 0\\
			0&-\varepsilon_{-\boldsymbol{k}h \bar{\eta}\uparrow} \\
		\end{array}
		\right) \left(
		\begin{array}{cc}
			\alpha_{\boldsymbol{k}e \eta\uparrow}\\
			\alpha^{\dagger}_{-\boldsymbol{k}h \bar{\eta}\uparrow} \\
		\end{array}
		\right),
	\end{equation}
\end{small}where we set $\varepsilon_{\boldsymbol{k}e \eta\uparrow}=\varepsilon_{\boldsymbol{k}\boldsymbol{+} \eta\uparrow}$ and $-\varepsilon_{-\boldsymbol{k}h \bar{\eta}\uparrow}=\varepsilon_{\boldsymbol{k}\boldsymbol{-} \eta\downarrow}$. Notably, the nonequal chemical potentials $M_e$ and $-M_h$ of the electron band $\varepsilon_{\boldsymbol{k}e \eta\uparrow}=\varepsilon_{\boldsymbol{k}}-M_e$ and hole band $\varepsilon_{\boldsymbol{k}h \bar{\eta}\uparrow}=\varepsilon_{\boldsymbol{k}}+M_h$ result in a splitted Fermi surface of the electron and hole states, which can be effectively viewed as a Zeeman split induced by the ``magnetic field''.

The interaction part $H_{\mathrm{int}}$ includes terms given by $\alpha^{\dagger}_{\boldsymbol{k}-\boldsymbol{q}e \eta\uparrow} \alpha_{-(\boldsymbol{k'}+\boldsymbol{q})h \bar{\eta}\uparrow} \alpha^{\dagger}_{-\boldsymbol{k}'h \bar{\eta}\uparrow}  \alpha_{\boldsymbol{k}e \eta\uparrow}$. In the presence of the splitted Fermi surface, an exotic Fulde-Ferrell-Larkin-Ovchinnikov superconducting state with nonzero momentum paring was predicted to appear\cite{FFstate1964,FFLOPerMag2006,FFLOHeaFer2007,FFLOultr2018,FFLOFeSe2020}. Whereas, its formation requires large split of the spin-resolved chemical potentials, or the BCS state with zero momentum paring will continue to dominate\cite{FFLOPerMag2006,FFLOHeaFer2007,FFLOultr2018}. In this work, the imbalance of $M_e$ and $-M_h$ is small with the chosen parameters. Thus, we can reasonably reserve terms satisfying $\boldsymbol{k}'+\boldsymbol{q}=\boldsymbol{k}$ in $H_{\mathrm{int}}$, which pair electron and hole with zero momentum ($-\boldsymbol{k}h \bar{\eta}\uparrow, \boldsymbol{k}e \eta\uparrow$). Under BCS paring, $H_{\mathrm{int}}$ transforms to the attractive e-h interaction
\begin{equation}
	H_{\mathrm{int}}= -\sum_{\eta,\boldsymbol{k},\boldsymbol{k}'} U^{\eta}_{\boldsymbol{k}\boldsymbol{k}'}  \alpha^{\dagger}_{\boldsymbol{k}'e \eta\uparrow}  \alpha^{\dagger}_{-\boldsymbol{k}'h \bar{\eta}\uparrow}  \alpha_{-\boldsymbol{k}h \bar{\eta}\uparrow}  \alpha_{\boldsymbol{k}e \eta\uparrow},
\end{equation}
where the interaction matrix element
\begin{equation}
  U^{\eta}_{\boldsymbol{k}\boldsymbol{k}'}
   = \frac{1}{N} \sum_{\alpha,\beta} \left[M^{\alpha,\ast}_{\boldsymbol{k}'\boldsymbol{+} \eta\uparrow} M^{\beta,\ast}_{\boldsymbol{k}\boldsymbol{-} \eta\downarrow}    M^{\beta}_{\boldsymbol{k}'\boldsymbol{-} \eta\downarrow} M^{\alpha}_{\boldsymbol{k}\boldsymbol{+} \eta\uparrow} \right] V^{\alpha\beta}_{\boldsymbol{k}\boldsymbol{k}'},
\end{equation}
$V^{\alpha\beta}_{\boldsymbol{k}\boldsymbol{k}'}=\sum_{j} V^{\alpha\beta}_{0j} e^{-i(\boldsymbol{k}-\boldsymbol{k}')(\boldsymbol{R}_j+\boldsymbol{\delta}_\beta-\boldsymbol{\delta}_\alpha)}$ with the nonlocal real-space Coulomb interaction matrix element $ V^{\alpha\beta}_{0j}=\langle\alpha0|\langle\beta j|V(\boldsymbol{r}-\boldsymbol{r}') |\beta j\rangle |\alpha0\rangle= \langle\alpha0|\langle\beta j| \frac{g\hbar\nu_{F}}{|\boldsymbol{r}-\boldsymbol{r}'|} |\beta j\rangle |\alpha0\rangle$, and $M^{\alpha(\beta)}_{\boldsymbol{k} \boldsymbol{\pm} \eta\sigma}$ is the bare system eigenstate of spin $\sigma$($\uparrow,\downarrow$) band in valley $\boldsymbol{K}_{\eta}$ with momentum $\boldsymbol{k}$, projected on sublattice $\alpha$ $(\beta)$. The mean-field approximation of $H_{\mathrm{int}}$ is
\begin{equation}\label{HintSC}
	H_{\mathrm{int}}\approx - \sum_{\eta,\boldsymbol{k}} \Delta^{\eta}_{\boldsymbol{k}} \alpha^{\dagger}_{\boldsymbol{k}e \eta\uparrow}  \alpha^{\dagger}_{-\boldsymbol{k}h \bar{\eta}\uparrow} +
	\Delta^{\eta,\ast}_{\boldsymbol{k}} \alpha_{-\boldsymbol{k}h \bar{\eta}\uparrow}  \alpha_{\boldsymbol{k}e \eta\uparrow},
\end{equation}
where the exciton pair potential is defined as $\Delta^{\eta}_{\boldsymbol{k}}=\sum_{\boldsymbol{k}'} U^{\eta}_{\boldsymbol{k}'\boldsymbol{k}} \langle \alpha_{-\boldsymbol{k}'h \bar{\eta}\uparrow}  \alpha_{\boldsymbol{k}'e \eta\uparrow} \rangle$.

Combing Eqs. \eqref{H0SC} and \eqref{HintSC}, we have the total mean-field Hamiltonian $H$ in the BdG representation
\begin{small}
\begin{equation}\label{Hbdg}
	H=\sum_{\eta,\boldsymbol{k}} 
	\left(
	\begin{array}{cc}
		\alpha^{\dagger}_{\boldsymbol{k}e \eta\uparrow}&\alpha_{-\boldsymbol{k}h \bar{\eta}\uparrow}\\
	\end{array}
	\right)
	\left(
	\begin{array}{cc}
		\varepsilon_{\boldsymbol{k}e \eta\uparrow} & -\Delta^{\eta}_{\boldsymbol{k}}\\
		-\Delta^{\eta,\ast}_{\boldsymbol{k}}&-\varepsilon_{-\boldsymbol{k}h \bar{\eta}\uparrow}\\
	\end{array}
	\right)
	\left(
	\begin{array}{cc}
		\alpha_{\boldsymbol{k}e \eta\uparrow}\\
		\alpha^{\dagger}_{-\boldsymbol{k}h \bar{\eta}\uparrow}\\
	\end{array}
	\right).
\end{equation} 
\end{small}Since the two valleys $\boldsymbol{K}_{\pm}$ are degenerate, we only need to deal with one single valley and omit the index $\eta$ afterwards. When $M_e=-M_h=M$, the energy spectrum of $H_{\rm BdG}$ are given by $E_{\boldsymbol{k}\boldsymbol{\pm}}=\pm\sqrt{(\varepsilon_{\boldsymbol{k}}-M)^2+|\Delta_{\boldsymbol{k}}|^2}$, which are schematically plotted in Fig. \ref{fig4}(c). Compared to the bands in the absence of pair potential [see dashed lines in Fig. \ref{fig4}(c)], an energy gap $\Delta$ is opened.
This means that when an electron and a hole attract and bind into e-h pair, the energy of the system is lowered.
Therefore, the ground state of the FM phased ABCA graphene is a superfluid composed of spin-triplet exciton pairs($-\boldsymbol{k}h\uparrow, \boldsymbol{k}e\uparrow$), namely, a spin SC. Analogous to the normal SC, a spin supercurrent can flow without dissipation in the spin SC. On the other hand, because of the neutral charge of the exciton pairs and the opened energy gap, no charge current can flow making the spin SC a charge insulator\cite{spinSCFMgra2011}. 

In spin SCs, the spin superconducting gap $\Delta(T)$ can be estimated by solving the self-consistent equation\cite{FFLOultr2018}:
\begin{equation}\label{gapeq}
	\Delta_{\boldsymbol{k}}= \sum_{\boldsymbol{k}'} \frac{U_{\boldsymbol{k}'\boldsymbol{k}}\Delta_{\boldsymbol{k}'}}{2E_{\boldsymbol{k}'}} [f(E_{\boldsymbol{k}'\boldsymbol{-}})- f(E_{\boldsymbol{k}'\boldsymbol{+}})],
\end{equation}
where $E_{\boldsymbol{k}\boldsymbol{\pm}}=-(M_e+M_h)/2\pm E_{\boldsymbol{k}}$ with $E_{\boldsymbol{k}}=\sqrt{[\varepsilon_{\boldsymbol{k}}-(M_e-M_h)/2]^2+|\Delta_{\boldsymbol{k}}|^2}$, $f(E)=1/(e^{E/K_B T}+1)$ is the Fermi-Dirac distribution function with $T$ the temperature.
Due to that $U_{\boldsymbol{k}'\boldsymbol{k}}$ varies slowly with $\boldsymbol{k}'-\boldsymbol{k}$, let us assume $U_{\boldsymbol{k}'\boldsymbol{k}}\approx U_{\boldsymbol{k}\boldsymbol{k}}\theta(k_D-|\boldsymbol{k}'-\boldsymbol{k}|) =\frac{U_0}{N} \theta(k_D-|\boldsymbol{k}'-\boldsymbol{k}|)$, where $k_D$ is the momentum cut-off and
\begin{equation}
	U_{\boldsymbol{k}\boldsymbol{k}}=\frac{U_0}{N}=\sum_{\alpha,\beta} \left[M^{\alpha,\ast}_{\boldsymbol{k}\boldsymbol{+} \eta\uparrow} M^{\beta,\ast}_{\boldsymbol{k}\boldsymbol{-} \eta\downarrow} M^{\beta}_{\boldsymbol{k}\boldsymbol{-} \eta\downarrow} M^{\alpha}_{\boldsymbol{k}\boldsymbol{+} \eta\uparrow} \right] \sum_{j} \frac{V^{\alpha\beta}_{0j}}{N}.
\end{equation}
Note that $V^{\alpha\beta}_{0j}$ depends linearly on the parameter $g$, so does $U_0$.
Invoking the values of real-space nonlocal interaction $ V^{\alpha\beta}_{0j}$ given by Ref. \cite{StrCouInt2011}, we thus can estimate $U_0\approx 28.0$ eV in ABCA graphene (referring to $g=2.1$).
Under the above assumption, Eq. \eqref{gapeq} is reduced to
\begin{equation}\label{gapzero}
	1=U_0\int_{0}^{\varepsilon_D} d\varepsilon ~ \frac{\rho(\varepsilon)[f(E_{\varepsilon\boldsymbol{-}})- f(E_{\varepsilon\boldsymbol{+}})]}{\sqrt{(2\varepsilon-M_e +M_h)^2+4\Delta(T)^2}}
\end{equation}
where $\rho(\varepsilon)\sim 1/\sqrt{\varepsilon}$ is the DOS of bare ABCA graphene for one valley and one spin channel, and the energy cut-off $\varepsilon_D$ is a function of $k_D$. 
Due to that $M_e$ and $M_h$ are equilibrium properties that are determined by doping $n$ and interaction $g$, $\Delta(T)$ should also depend on them according to Eq. \eqref{gapzero}.
We numerically calculate $\Delta$ as a function of $T$ for several $n$ and $g$, with $\varepsilon_D=\gamma$. As shown in Fig. \ref{fig4}(c), $\Delta$ decreases with $T$ as expected. At $T=0$ K, it is less affected by $n$ with a gap $\Delta\approx 7.0$ meV for $g=2.1$ (see solid lines), which is about 2 times larger than that in FM graphene ($\approx 3$ meV \cite{spinSCFMgra2011}) and comparable to that in trilayer graphene ($\approx 7.8$ meV \cite{spinsctri2012}). At $T\neq0$ K, $\Delta$ and thus the spin superconductivity is suppressed by enlarging doping $n$, as the Fermi surface split and equivalent ``magnetic filed'' grow larger driving the system out of the spin SC state. In screened system with $g=1.9$, the spin split-energy is decreased, thus giving rise to smaller gaps $\Delta$ compared to that given by $g=2.1$ (see dash lines).

The critical temperature $T_c$ can be estimated by bring $\Delta(T_c)=0$ into Eq. \eqref{gapzero}.
We numerically calculate $T_c$ as a function of $n$ for different $g$, with $\varepsilon_D=\gamma$. As shown in Fig. \ref{fig4}(d), $T_c$ is lowered by increasing $n$, which is attributed to the effect of the equivalent ``magnetic filed'' as explained above. It can as well be suppressed by decreasing $g$ from 2.1 to 1.9, as illustrated by the solid and dash lines. For non-doping system with $n=0$ and $g=2.1$, $T_c$ reaches about $45$ K.
The spin superconducting gap $\Delta(0)\approx 7.0$ meV and critical temperature $T_c\approx 45$ K is moderate, which should make the spin SC state in ABCA graphene readily to confirm under the state-of-the-art experimental conditions.

One more note to stress, the spin SC state in ABCA graphene can not exist if the system is heavily doped with large $n$, where FM phase disappears [see Fig. \ref{fig3}(c)].

\section{Spin-current Josephson effect}\label{spinJoe}
\begin{figure}
	\centering
	\includegraphics[width=1.0\columnwidth]{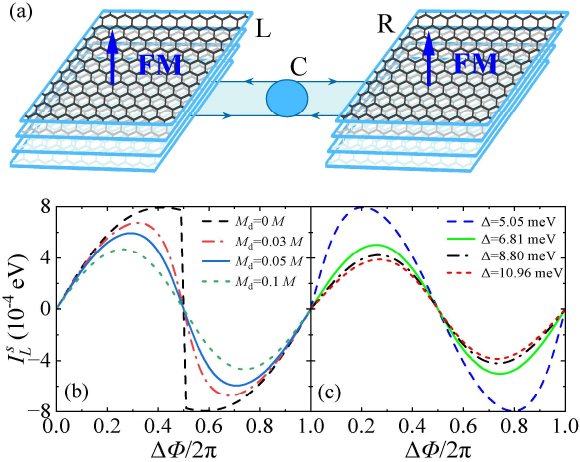}
	\caption{(Color online) (a) Schematic of the spin SC/quantum dot/spin SC heterojunction. L/R labels the left/right FM-phased ABCA graphene leads and C is the central quantum dot. Blue arrows denote magnetization direction. (b) Spin supercurrent $I^{s}_{L}$ versus phase difference $\Delta\phi/2\pi$ for different $M_d$. Here we set $M=0.036\gamma$, $\Delta=7$ meV ($g=2.1$ and $n=0$), $\Gamma=0.05 M$ and $\epsilon_d=0$. (c) $I^{s}_{L}$ versus $\Delta\phi/2\pi$ for different $\Delta$. Here we choose $n=0$ and $g=1.9, 2.1, 2.3,2.5$, which gives $M=(0.0356,0.0364,0.0375,0.0381)\gamma$ and $\Delta=(5.05,6.81,8.80,10.96)$ meV. Other parameters are $\epsilon_d=0$, $M_d=0.1 M$ and $\Gamma=0.05 M$.}
	\label{fig5}
\end{figure}

Josephson effect is a fundamental hallmark of the charge SCs and has widely applications. As a counterpart of the SC, the spin SC holds a similar spin-current Josephson effect,
which has been demonstrated in the FM graphene \cite{spinSCFMgra2011}. Here, we will study it in a junction composed of FM ABCA graphene spin SC /quantum dot/FM ABCA graphene spin SC, as shown in Fig. \ref{fig5}(a). The junction Hamiltonian comprises three part: $H=\sum_{\beta=L,R}H_{\beta}+H_{\mathrm{dot}}+H_{\mathrm{coup}}$, where the lead Hamiltonian $H_{\beta}$, the quantum dot $H_{\mathrm{dot}}$ and their coupling $H_{\mathrm{coup}}$ are, respectively,
\begin{small}
\begin{equation}
	\begin{split}
		& H_{\beta} = \sum_{\eta,\boldsymbol{k}}
		\left(
		\begin{array}{cc}
			\alpha^{\dagger}_{\beta\boldsymbol{k}\boldsymbol{+} \eta\uparrow}&\alpha^{\dagger}_{\beta\boldsymbol{k}\boldsymbol{-} \eta\downarrow}\\
		\end{array}
		\right)
		\left(
		\begin{array}{cc}
			\varepsilon_{\boldsymbol{k}}-M & \Delta e^{i\phi_\beta}\\
			\Delta e^{-i\phi_\beta}&-\varepsilon_{\boldsymbol{k}}+M\\
		\end{array}
		\right)
		\left(
		\begin{array}{cc}
			\alpha_{\beta\boldsymbol{k}\boldsymbol{+} \eta\uparrow}\\
			\alpha_{\beta\boldsymbol{k}\boldsymbol{-} \eta\downarrow}\\
		\end{array}
		\right), \\
		& H_{\mathrm{dot}}= \sum_{\sigma=\uparrow,\downarrow}(\epsilon_d+\sigma M_d) d^{\dagger}_{\sigma}d_{\sigma}, \\
		& H_{\mathrm{coup}}= \sum_{\beta,\eta,\boldsymbol{k}} (t_\beta \alpha^{\dagger}_{\beta\boldsymbol{k}\boldsymbol{+} \eta\uparrow}d_{\uparrow}+ t_\beta \alpha^{\dagger}_{\beta\boldsymbol{k}\boldsymbol{-} \eta\downarrow}d_{\downarrow} + \mathrm{H.c.}).
	\end{split}
\end{equation}
\end{small}Here, $H_{\beta}$ is the same as $H_{\mathrm{BdG}}$ in Eq. \eqref{Hbdg}. $\beta=L/R$ denotes the clean spin SC lead ($n=0$) with $M_e=-M_h=M$, and its order parameter $\Delta e^{i\phi_\beta}$ ($\Delta$ and $\phi_\beta$ are SC gap and phase, respectively) is assumed to be independent of momentum $\boldsymbol{k}$. $\epsilon_d$ is the quantum dot level with a spin-split energy $M_d$. $t_\beta$ is the coupling of the lead with the quantum dot.

Using the non-equilibrium Green's function method, the spin-resolved current flowing from lead $\beta$ to the central region at equilibrium can be calculated by $I_{\beta\uparrow(\downarrow)}= \mathrm{Re} \frac{2et_\beta}{\hbar} \sum_{\eta,\boldsymbol{k}} G^{<}_{d\beta\boldsymbol{k}\eta,11(22)}(t,t)$, where $G^{<}_{d\beta\boldsymbol{k}\eta,11(22)}(t,t)$ is the up (down) diagonal element of the lesser Green's function matrix defined in the Nambu basis.
Applying Dyson's equation to expand $G^{<}_{d\beta\boldsymbol{k}\eta}(t,t)$, the spin-resolved currents flowing in the left lead can then be derived as (details can be found in Appendix \ref{appdspcur})
\begin{equation}\label{Lspincurr}
	\begin{split}
		&I_{L\uparrow}\approx  \frac{4e}{\hbar} \int \frac{d\varepsilon}{2\pi} ~ \frac{\Gamma^2\Delta^2}{\varepsilon^2-\Delta^2}f(\varepsilon) \mathrm{Im} \frac{1}{B^{\ast}(\varepsilon) }\mathrm{sin}(\phi_R-\phi_L); \\
		&I_{L\downarrow}\approx  \frac{4e}{\hbar} \int \frac{d\varepsilon}{2\pi} ~ \frac{\Gamma^2\Delta^2}{\varepsilon^2-\Delta^2}f(\varepsilon) \mathrm{Im} \frac{1}{B^{\ast}(\varepsilon) }\mathrm{sin}(\phi_L-\phi_R),
	\end{split}
\end{equation}
where $\Gamma=2\pi |t_\beta|^2 \rho_N(0)$ is the linewidth function of $\beta$ lead with $\rho_N(0)$ the DOS of the lead in norm state ($\Delta=0$) at Fermi energy $E_f=0$. The expression of $B(\varepsilon)$ is complex and thus shown in the appendix.

Above equations reveal that the tunneling current with spin up is opposite to that with spin down, rendering a vanishing charge current  $I^{e}_{L}= (I_{L\uparrow}+I_{L\downarrow})=0$. This is in agreement with the fact that the charge neutral e-h pairs in the spin SC can not transport charge current. Whereas, there is a net spin supercurrent flowing through the junction that resembles the Josephson effect in the conventional SCs:
\begin{equation}
\begin{split}
	I^{s}_{L}=& -(I_{L\uparrow}-I_{L\downarrow}) \frac{\hbar}{2e} \\ \approx& \int \frac{d\varepsilon}{2\pi} ~ 
	\frac{4\Gamma^2\Delta^2}{\varepsilon^2-\Delta^2} f(\varepsilon) \mathrm{Im} \frac{1}{B^{\ast}(\varepsilon)} \mathrm{sin}(\phi_L-\phi_R).
\end{split}
\end{equation}
We numerically calculate $I^{s}_{L}$ as functions of phase difference $\Delta \phi=\phi_L-\phi_R$ for several $M_d$ and $\Delta$, which are shown in Figs. \ref{fig5}(b-c).
The spin supercurrent $I^{s}_{L}$ can exist with non-zero $\Delta \phi$ and $\Delta$, even in a spin degenerate quantum dot where $M_d=0$.
In summary, we have shown that the junction composed of ABCA graphene spin SCs and quantum dot harbors a net spin supercurrent at equilibrium, manifesting the spin-current Josephson effect.

\section{Conclusions}\label{conclu}

In conclusion, we demonstrate that the ground state of ABCA-stacked tetralayer graphene can be a spontaneous spin SC state at low temperature. This finding is quantitatively elucidated by first studying the FM exchange instability and the equilibrium phase of the system in the presence of long-range e-e interaction. The phase diagram indicates that at non- or low-doping levels, the interaction will stabilize a FM phase with the coexistence of electron and hole pockets in the two spin channels, respectively. However, this FM state becomes unstable when considering the naturally occurring attractive e-h interaction. As a result, electrons and holes spontaneously form spin-triplet excitonic pairs, which further condense into a spin superfluid, termed spin SC. We developed a consistent BCS-type theory for this spin SC state in ABCA graphene, estimating the superconducting gap and critical temperature to be approximately 7.0 meV and 45 K for the non-doping system. Tetralayer graphene, superior to bilayer and trilayer forms, offers more flexibility in manipulating the FM phase, potentially enhancing the realization of the spin SC state. Additionally, a spin-current Josephson effect is demonstrated in the ABCA graphene spin SC/quantum dot/ABCA graphene spin SC junction.

\begin{acknowledgments}
We thank Y.-H. Li, M. Gong and H.-L. Li for fruitful discussions. This work is financially supported by the National Key R\&D Program of China (Grants No. 2019YFA0308403), and the National Natural Science Foundation of China (Grant No. 12350401).

S. Li and Y.-H. Ren contributed equally to this work.
\end{acknowledgments}

\onecolumngrid
\noindent
\begin{appendices}
\renewcommand{\theequation}{S\arabic{equation}}
\renewcommand{\thefigure}{S\arabic{figure}}
\setcounter{figure}{0}
\setcounter{equation}{0}
	
\appendix

\section{Derivation of the e-e exchange interaction}\label{appdex}
Here we provide the derivation of Eq. \eqref{chardensHex} in the main text starting from the second quantization formula Eq. \eqref{secHex}.
First, we expand the field operators $\varphi(\boldsymbol{r})$ in the orbital Bloch wave function basis described by $\psi_{\alpha \eta\sigma}(\boldsymbol{k})$
\begin{equation}
	\varphi^{\dagger}(\boldsymbol{r})=\sum_{\alpha\boldsymbol{k}\sigma} \psi^{\dagger}_{\alpha\sigma}(\boldsymbol{k}) ~ \langle \alpha\boldsymbol{k}\sigma|; ~~~
	\varphi(\boldsymbol{r})=\sum_{\alpha\boldsymbol{k}\sigma}  |\alpha\boldsymbol{k}\sigma \rangle ~ \psi_{\alpha\sigma}(\boldsymbol{k}),
\end{equation}
where $\alpha (\mathrm{A}_1,\cdots,\mathrm{B}_4)$ is the sublattice index. The valley index $\eta$ has been omitted for that momentum $\boldsymbol{k}$ is summed over the whole BZ. Bringing above into Eq. \eqref{secHex}, we have
\begin{equation}
	H_{\rm ex}= \frac{1}{2} \sum_{\alpha_1\boldsymbol{k}_1\sigma_1} \cdots \sum_{\alpha_4\boldsymbol{k}_4\sigma_4} 
	\langle \alpha_1\boldsymbol{k}_1\sigma_1,\alpha_2\boldsymbol{k}_2\sigma_2| V(\boldsymbol{r}-\boldsymbol{r}') | \alpha_3\boldsymbol{k}_3\sigma_3,\alpha_4\boldsymbol{k}_4\sigma_4 \rangle
	\psi^{\dagger}_{\alpha_1\sigma_1}(\boldsymbol{k}_1) \psi^{\dagger}_{\alpha_2\sigma_2}(\boldsymbol{k}_2)  \psi_{\alpha_3\sigma_3}(\boldsymbol{k}_3) \psi_{\alpha_4\sigma_4}(\boldsymbol{k}_4),
\end{equation}
where the Coulomb interaction $V(\boldsymbol{r}-\boldsymbol{r}')=\frac{e^2}{\epsilon|\boldsymbol{r}-\boldsymbol{r}'|}$. The interaction matrix elements can be simplified a step further. Suppose that the sublattices $\alpha_1$ and $\alpha_4$ belong to layer $j$, and that $\alpha_2$ and $\alpha_3$ locate on layer $i$, then
\begin{equation}
	\begin{split}
		\langle \alpha_1\boldsymbol{k}_1\sigma_1,\alpha_2\boldsymbol{k}_2\sigma_2| V(\boldsymbol{r}_i-\boldsymbol{r}_j)  |\alpha_3\boldsymbol{k}_3\sigma_3,\alpha_4\boldsymbol{k}_4\sigma_4 \rangle
		&=\frac{\delta_{\sigma_1,\sigma_4}\delta_{\sigma_2,\sigma_3}}{S} \sum_{\boldsymbol{q} \neq \boldsymbol{0}} V_{ij}(\boldsymbol{q})
		\langle \alpha_1\boldsymbol{k}_1,\alpha_2\boldsymbol{k}_2| e^{i\boldsymbol{q}\cdot(\boldsymbol{r}_i-\boldsymbol{r}_j)} | \alpha_3\boldsymbol{k}_3,\alpha_4\boldsymbol{k}_4 \rangle \\
		&= \frac{\delta_{\sigma_1,\sigma_4}\delta_{\sigma_2,\sigma_3} \delta_{\alpha_1,\alpha_4}\delta_{\alpha_2,\alpha_3}}{S} \sum_{\boldsymbol{q}\neq\boldsymbol{0}}
		\sum_{\boldsymbol{G,G'}}
		V_{ij}(\boldsymbol{q}) \delta_{\boldsymbol{k}_3+\boldsymbol{q},\boldsymbol{k}_2+\boldsymbol{G}} \delta_{\boldsymbol{k}_4-\boldsymbol{q},\boldsymbol{k}_1+\boldsymbol{G'}},
	\end{split}
\end{equation}
where $S$ denotes the area of the system, $V_{ij}(\boldsymbol{q})=\frac{2\pi e^2}{\epsilon q} e^{-qd_{ij}}$ is the Fourier component of $V(\boldsymbol{r}_i-\boldsymbol{r}_j)$ with $d_{ij}$ the vertical distance between layers $i$ and $j$, $\boldsymbol{G}$ and $\boldsymbol{G}'$ are reciprocal lattice vectors and $\delta$'s are the Kronecker symbols.
Here, the $\boldsymbol{q}=\boldsymbol{0}$ term is excluded due to the compensation of interaction from the positive ionic background. To obtain the second line of above equation, we have invoked the relation that
\begin{equation}
	\langle \alpha\boldsymbol{k}_1| e^{i\boldsymbol{q}\cdot \boldsymbol{r}} | \beta\boldsymbol{k}_2 \rangle=\frac{1}{N} \sum_{i,j} e^{i\boldsymbol{k}_2\cdot \boldsymbol{R}^{\beta}_{j}}
	e^{-i\boldsymbol{k}_1\cdot \boldsymbol{R}^{\alpha}_{i}}  \langle \alpha i| e^{i\boldsymbol{q}\cdot \boldsymbol{r}} | \beta j \rangle
	=\frac{\delta_{\alpha,\beta}}{N} \sum_i  e^{i(\boldsymbol{k}_2+\boldsymbol{q}-\boldsymbol{k}_1) \cdot \boldsymbol{R}^{\alpha}_{i}}
	=\sum_{\boldsymbol{G}} \delta_{\alpha,\beta} \delta_{\boldsymbol{k}_2+\boldsymbol{q},\boldsymbol{k}_1+\boldsymbol{G}},
\end{equation}
where $R^{\alpha}_{i}$ ($R^{\beta}_{j}$) represents the lattice vector of sublattice $\alpha$ ($\beta$) in the $i$th ($j$th) unit cell.
Therefore in $H_{\mathrm{ex}}$, the contribution from layers $i$ and $j$ can be recast to
\begin{equation}
	\begin{split}
		H^{ji}_{\mathrm{ex}}= \frac{1}{2S}\sum_{\alpha_1\epsilon j}\sum_{\alpha_2\epsilon i} \sum_{\boldsymbol{k}_1,\boldsymbol{k}_2}\sum_{\boldsymbol{q}\neq\boldsymbol{0}} \sum_{\sigma}
		V_{ij}(\boldsymbol{q}) ~[\psi^{\dagger}_{\alpha_1\sigma}(\boldsymbol{k}_1) \psi^{\dagger}_{\alpha_2\sigma}(\boldsymbol{k}_2)  &
		\psi_{\alpha_2\sigma}(\boldsymbol{k}_2-\boldsymbol{q}) \psi_{\alpha_1\sigma}(\boldsymbol{k}_1+\boldsymbol{q})  +  \\
		& \psi^{\dagger}_{\alpha_1\sigma}(\boldsymbol{k}_1) \psi^{\dagger}_{\alpha_2\bar{\sigma}}(\boldsymbol{k}_2) \psi_{\alpha_2\bar{\sigma}}(\boldsymbol{k}_2-\boldsymbol{q}) \psi_{\alpha_1\sigma}(\boldsymbol{k}_1+\boldsymbol{q})].
	\end{split}
\end{equation}
We can define the Fourier component of the layer-dependent charge density operator $\hat{\rho}_i(\boldsymbol{r}_i)$ as usual: $\hat{\rho}_{i}(\boldsymbol{q})=\sum_{\alpha\boldsymbol{p}\sigma}  
\psi^{\dagger}_{\alpha_i\sigma}(\boldsymbol{p}+\boldsymbol{q}) \psi_{\alpha_i\sigma}(\boldsymbol{p})$.
By summing all the contributions from different combinations of $i$ and $j$, the exchange interaction can be written in the form that
\begin{equation}\label{chardensHcoul}
	H_{\mathrm{ex}}=\sum_{i,j} H^{ji}_{\mathrm{ex}}= \frac{1}{2S}\sum_{\boldsymbol{q}\neq\boldsymbol{0}}\sum_{i,j} \hat{\rho}_{i}(\boldsymbol{q})  V_{ij}(\boldsymbol{q}) \hat{\rho}_{j}(-\boldsymbol{q}),
\end{equation}
which is the expression shown in Eq. \eqref{chardensHex} of the main text.

\section{Equilibrium properties of non-doping ABCA graphene}\label{appendhf}
\begin{figure}[htb]
	\centering
	\includegraphics[width=0.7\textwidth]{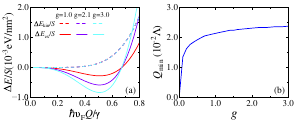}
	\caption{(Color online) (a) $\Delta E_{\mathrm{kin}}/S$ and $\Delta E_{\mathrm{ex}}/S$ of the FM variational state as functions of variational parameter $Q$ for various $g$. (b) The equilibrium pocket $Q_{\mathrm{min}}$ vs $g$ for half filling ABCA graphene ($n=0$).}
	\label{sfig1}
\end{figure}

Here, we give a complementary discussion of the equilibrium properties of non-doping ABCA graphene in the presence of e-e interaction. First, we investigate the effects of $g$. As shown in Fig. \ref{sfig1}(a), we numerically calculate the kinetic $\Delta E_{\mathrm{kin}}/S$ and exchange $\Delta E_{\mathrm{ex}}/S$ energy densities of the FM variational states versus $Q$ for $g=1.0,2.1$ and $3.0$. $\Delta E_{\mathrm{kin}}/S$ is invariant with $g$ for sure because it is not relevant to interactions. Whereas, $\Delta E_{\mathrm{ex}}/S$ is firstly lowered with the 
increase of $g$, and then goes up by increasing $g$ when $Q$ exceeds some critical value. The position of the minimum of $\Delta E_{\mathrm{ex}}/S$ is affected by $g$, thus rendering the equilibrium pocket $Q_{\mathrm{min}}$ dependent on $g$ as well. We show the $Q_{\mathrm{min}}$-$g$ relation in Fig. \ref{sfig1}(b). $Q_{\mathrm{min}}$ first goes up rapidly and then saturates for large enough $g$. The above study reveals that the FM phase is more stabilized in the half filling system with increasing $g$.

\begin{figure}[htb]
	\centering
	\includegraphics[width=0.7\textwidth]{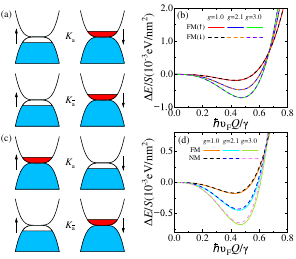}
	\caption{(Color online) (a) The spin-flipped variational state with respect to that given in Fig. \ref{fig2}(a). (b) $\Delta E/S$ vs $Q$ for the spin-flipped state with $\downarrow$ magnetization (dash lines) and for the FM state with $\uparrow$ magnetization (solid lines). (c) Nonmagnetic variational state with reversed pockets between the two valleys $\boldsymbol{K}_a$ and $\boldsymbol{K}_{\bar{a}}$, and between the two spin channels. (d)  $\Delta E/S$ vs $Q$ for the NM state (dash lines) and for the FM stat (solid lines), incorporating intervalley scattering contribution.}
	\label{sfig2}
\end{figure}

Second, we discuss the magnetization direction of the equilibrium FM phase. Consider a spin-flipped variational state where the pockets of each spin channel are opposite to that in Fig. \ref{fig2}(a), as shown in Fig. \ref{sfig2}(a). We then calculate $\Delta E/S$ vs $Q$ for this state with $\downarrow$ magnetization to compare with the FM state with $\uparrow$ magnetization [see Fig. \ref{sfig2}(b)].  A clear degeneracy in energy is found between the two cases, demonstrating a degenerate FM state at equilibrium with reversed magnetization. In another words, FM equilibrium states have no spin preference.

Furthermore, we study the NM variational state where the pockets are reversed at the two valleys and also opposite in the two spin channels, as displayed in Fig. \ref{sfig2}(c). The FM state preserves $\mathrm{Z}_2$ symmetry but with broken $\mathrm{SU}(2)$ symmetry. Whereas, the NM state breaks both. We then calculate their energy density $\Delta E/S$ vs $Q$ by taking intervalley scattering contribution into account, to see which is more energetically favorable. The exchange energy density contributed by intervalley scattering is given by
\begin{equation}
	\frac{E^{\mathrm{inter}}_{\mathrm{ex}}}{S} = -\frac{1}{2} \int_D \frac{d^2\boldsymbol{p}}{(2\pi)^2} \int_D \frac{d^2\boldsymbol{q}}{(2\pi)^2} \sum_{i,j} \sum_{m,n,\sigma} \sum_{a\neq a'} 
	\chi^{i}_{mn,a'a}(\boldsymbol{q},\boldsymbol{p}) \chi^{j}_{nm,aa'}(\boldsymbol{p},\boldsymbol{q}) V_{ij}(\boldsymbol{p}-\boldsymbol{q}+\boldsymbol{K}_a-\boldsymbol{K}_{a'})  n_{ma'\sigma}(\boldsymbol{q}) n_{na\sigma}(\boldsymbol{p}).
\end{equation}
Put this together with the intravalley contribution Eq.\eqref{exden}, we can calculate the total $E_{\mathrm{ex}}/S$ and then $\Delta E/S$. As shown in Fig. \ref{sfig2}(d), $\Delta E/S$ of the FM state is lower than that of the NM state, manifesting that the FM state is more favorable and stable at equilibrium than the NM state. Notice that the energy differences between the two states are tiny, which is due to that the intervalley scattering contribution to the exchange energy $E^{\mathrm{inter}}_{\mathrm{ex}}$ is two or three orders of magnitudes smaller than $E^{\mathrm{intra}}_{\mathrm{ex}}$ contributed by intravalley scattering. The results given in Fig. \ref{sfig2}(d) are slightly deviated from that shown in Fig. \ref{sfig2}(b) without $E^{\mathrm{inter}}_{\mathrm{ex}}$ contribution, making the differences between the FM and NM states insignificant.

\section{Derivation of the spin supercurrent}\label{appdspcur}
We give here a detailed derivation of the spin supercurrent in the spin SC heterojunction shown in Fig. \ref{fig5}, following the general non-equilibrium Green's function method.
The equilibrium current flowing from lead $\beta$ to the central region is given by
\begin{equation}
	I_{\beta}(t)=-e \langle \dot{\hat{N}}_{\beta} \rangle  = \frac{ie}{\hbar} \langle [\hat{N}_{\beta},\bar{H}](t) \rangle ,
\end{equation}
where $\hat{N}_{\beta}=\sum_{\boldsymbol{k},\eta} \alpha^{\dagger}_{\beta\boldsymbol{k}+ \eta\uparrow}\alpha_{\beta\boldsymbol{k}+ \eta\uparrow} + \alpha^{\dagger}_{\beta\boldsymbol{k}- \eta\downarrow}\alpha_{\beta\boldsymbol{k}- \eta\downarrow}$ is particle number operator of the lead, and $\bar{H}=H-\mu \hat{N}$ is the thermodynamic potential of the junction with $\hat{N}$ the total particle number operator. The commutator in the above equation is calculated by $[\hat{N}_{\beta},\bar{H}]=\sum_{\boldsymbol{k},\eta} t_\beta \alpha^{\dagger}_{\beta\boldsymbol{k}+ \eta\uparrow}d_{\uparrow} + t_\beta \alpha^{\dagger}_{\beta\boldsymbol{k}- \eta\downarrow}d_{\downarrow}- t^{\ast}_{\beta} d^{\dagger}_{\uparrow}\alpha_{\beta\boldsymbol{k}+ \eta\uparrow} - t^{\ast}_{\beta} d^{\dagger}_{\downarrow}\alpha_{\beta\boldsymbol{k}- \eta\downarrow}$.
Then the current $I_{\beta}$ can be recast to
\begin{equation}
	\begin{split}
		I_{\beta}(t)=& \frac{ie}{\hbar}  \sum_{\boldsymbol{k},\eta}  t_{\beta} \langle \alpha^{\dagger}_{\beta\boldsymbol{k}+ \eta\uparrow}(t)d_{\uparrow}(t) \rangle - t^{\ast}_\beta \langle d^{\dagger}_{\uparrow}(t) \alpha_{\beta\boldsymbol{k}+ \eta\uparrow}(t) \rangle + t_{\beta} \langle \alpha^{\dagger}_{\beta\boldsymbol{k}- \eta\downarrow}(t) d_{\downarrow}(t) \rangle - t^{\ast}_\beta \langle d^{\dagger}_{\downarrow}(t) \alpha_{\beta\boldsymbol{k}- \eta\downarrow}(t) \rangle \\
		=& \frac{2e}{\hbar}  \mathrm{Re} \sum_{\boldsymbol{k},\eta} t_{\beta}i \langle \alpha^{\dagger}_{\beta\boldsymbol{k}+ \eta\uparrow}(t) d_{\uparrow}(t) \rangle +  t_{\beta}i \langle \alpha^{\dagger}_{\beta\boldsymbol{k}- \eta\downarrow}(t) d_{\downarrow}(t) \rangle \\
		=& \frac{2e}{\hbar}  \mathrm{Re} \sum_{\boldsymbol{k},\eta}  t_\beta G^{<}_{d\beta\boldsymbol{k}\eta,11}(t,t) +  t_\beta G^{<}_{d\beta\boldsymbol{k}\eta,22}(t,t).
	\end{split}
\end{equation}
We can extract the spin-resolved currents from above equation, which read
\begin{equation}\label{spincurr}
\begin{split}
	I_{\beta\uparrow}(t)=& \mathrm{Re} \frac{2et_\beta}{\hbar} \sum_{\boldsymbol{k},\eta} G^{<}_{d\beta\boldsymbol{k}\eta,11}(t,t);  \\
	I_{\beta\downarrow}(t)=& \mathrm{Re} \frac{2et_\beta}{\hbar} \sum_{\boldsymbol{k},\eta} G^{<}_{d\beta\boldsymbol{k}\eta,22}(t,t),
\end{split}
\end{equation}
where the lesser Green's functions are given as the diagonal elements of the Green's function matrix defined in the Nambu representation
\begin{equation}\label{lesserG}
	G^{<}_{d\beta\boldsymbol{k}\eta}(t,t')=i
	\left(
	\begin{array}{cc}
		\langle \alpha^{\dagger}_{\beta\boldsymbol{k}+ \eta\uparrow}(t')d_{\uparrow}(t) \rangle & \langle \alpha^{\dagger}_{\beta\boldsymbol{k}- \eta\downarrow}(t')d_{\uparrow}(t) \rangle\\
		\langle \alpha^{\dagger}_{\beta\boldsymbol{k}+ \eta\uparrow}(t')d_{\downarrow}(t) \rangle&\langle \alpha^{\dagger}_{\beta\boldsymbol{k}- \eta\downarrow}(t')d_{\downarrow}(t) \rangle\\
	\end{array}
	\right).
\end{equation}
Furthermore, we also define the lesser and advanced Green's functions in the Nambu basis for the lead $\beta$ without coupling to the quantum dot by
\begin{equation}
	\begin{split}
		g^{<}_{\beta\boldsymbol{k}\eta}(t,t')=& i
		\left(
		\begin{array}{cc}
			\langle \alpha^{\dagger}_{\beta\boldsymbol{k}+ \eta\uparrow}(t')\alpha_{\beta\boldsymbol{k}+ \eta\uparrow}(t) \rangle_0 &\langle \alpha^{\dagger}_{\beta\boldsymbol{k}- \eta\downarrow}(t')\alpha_{\beta\boldsymbol{k}+ \eta\uparrow}(t) \rangle_0\\
			\langle \alpha^{\dagger}_{\beta\boldsymbol{k}+ \eta\uparrow}(t')\alpha_{\beta\boldsymbol{k}- \eta\downarrow}(t) \rangle_0&\langle \alpha^{\dagger}_{\beta\boldsymbol{k}- \eta\downarrow}(t')\alpha_{\beta\boldsymbol{k}- \eta\downarrow}(t) \rangle_0\\
		\end{array}
		\right); \\
		g^{a}_{\beta\boldsymbol{k}\eta}(t,t')=& i\theta(t'-t)
		\left(
		\begin{array}{cc}
			\langle\{ \alpha_{\beta\boldsymbol{k}+ \eta\uparrow}(t),\alpha^{\dagger}_{\beta\boldsymbol{k}+ \eta\uparrow}(t') \}\rangle_0 &\langle\{ \alpha_{\beta\boldsymbol{k}+ \eta\uparrow}(t),\alpha^{\dagger}_{\beta\boldsymbol{k}- \eta\downarrow}(t') \}\rangle_0\\
			\langle\{ \alpha_{\beta\boldsymbol{k}- \eta\downarrow}(t),\alpha^{\dagger}_{\beta\boldsymbol{k}+ \eta\uparrow}(t') \}\rangle_0&\langle\{ \alpha_{\beta\boldsymbol{k}- \eta\downarrow}(t),\alpha^{\dagger}_{\beta\boldsymbol{k}- \eta\downarrow}(t') \}\rangle_0\\
		\end{array}
		\right),
	\end{split}
\end{equation}
and the lesser and retarded/advanced Green's functions for the quantum dot with coupling to the leads by
\begin{equation}
	\begin{split}
		G^{<}_{dd}(t,&t')= i
		\left(
		\begin{array}{cc}
			\langle d^{\dagger}_{\uparrow}(t')d_{\uparrow}(t) \rangle &\langle d^{\dagger}_{\downarrow}(t')d_{\uparrow}(t) \rangle\\
			\langle d^{\dagger}_{\uparrow}(t')d_{\downarrow}(t) \rangle&\langle d^{\dagger}_{\downarrow}(t')d_{\downarrow}(t) \rangle\\
		\end{array}
		\right); \\
		G^{r/a}_{dd}(t,&t')= \mp i\theta(\pm t \mp t')
		\left(
		\begin{array}{cc}
			\langle\{ d_{\uparrow}(t),d^{\dagger}_{\uparrow}(t') \}\rangle &\langle\{ d_{\uparrow}(t),d^{\dagger}_{\downarrow}(t') \}\rangle\\
			\langle\{ d_{\downarrow}(t),d^{\dagger}_{\uparrow}(t') \}\rangle&\langle\{ d_{\downarrow}(t),d^{\dagger}_{\downarrow}(t') \}\rangle\\
		\end{array}
		\right).
	\end{split}
\end{equation}

We next solve the diagonal lesser Green's functions in Eq. \eqref{spincurr} from the Dyson's equation:
\begin{equation}
  G^{<}= g^{<}+G^{r}\Sigma^{r}g^{<} + G^{r}\Sigma^{<}g^{a} + G^{<}\Sigma^{a}g^{a}.
\end{equation}
Its non-diagonal matrix can be written as
\begin{equation}\label{Glessmat}
  G^{<}_{d\beta\boldsymbol{k}\eta}(t,t')= \frac{1}{\hbar}\int dt_1 ~ G^{r}_{dd}(t,t_1) \Sigma^{r}_{d\beta\boldsymbol{k}\eta}  g^{<}_{\beta\boldsymbol{k}\eta}(t_1,t') +  G^{<}_{dd}(t,t_1) \Sigma^{a}_{d\beta\boldsymbol{k}\eta}  g^{a}_{\beta\boldsymbol{k}\eta}(t_1,t'),
\end{equation}
where the retarded, advanced and lesser self-energies are given by
\begin{equation}
	\Sigma^{r,a}_{d\beta\boldsymbol{k}\eta} = \left(
	\begin{array}{cc}
		t^{\ast}_{\beta}&0\\
		0&t^{\ast}_{\beta}\\
	\end{array}
	\right), ~~~
	\Sigma^{<}_{d\beta\boldsymbol{k}\eta} = \left(
	\begin{array}{cc}
		0&0\\
		0&0\\
	\end{array}
	\right).
\end{equation}
Therefore, we can obtain the diagonal lesser Green's functions from Eq. \eqref{Glessmat} as
\begin{equation}
\begin{split}
	G^{<}_{d\beta\boldsymbol{k}\eta,11}(t,t')= \frac{1}{\hbar}\int dt_1& ~ G^{r}_{dd,11}(t,t_1) \Sigma^{r}_{d\beta\boldsymbol{k}\eta,11}   g^{<}_{\beta\boldsymbol{k}\eta,11}(t_1,t') +  G^{r}_{dd,12}(t,t_1) \Sigma^{r}_{d\beta\boldsymbol{k}\eta,22}  g^{<}_{\beta\boldsymbol{k}\eta,21}(t_1,t')+\\
	&  G^{<}_{dd,11}(t,t_1) \Sigma^{a}_{d\beta\boldsymbol{k}\eta,11}  g^{a}_{\beta\boldsymbol{k}\eta,11}(t_1,t') + G^{<}_{dd,12}(t,t_1) \Sigma^{a}_{d\beta\boldsymbol{k}\eta,22}  g^{a}_{\beta\boldsymbol{k}\eta,21}(t_1,t'); \\
	G^{<}_{d\beta\boldsymbol{k}\eta,22}(t,t')= \frac{1}{\hbar}\int dt_1& ~ G^{r}_{dd,22}(t,t_1) \Sigma^{r}_{d\beta\boldsymbol{k}\eta,22}  g^{<}_{\beta\boldsymbol{k}\eta,22}(t_1,t') +  G^{r}_{dd,21}(t,t_1) \Sigma^{r}_{d\beta\boldsymbol{k}\eta,11}  g^{<}_{\beta\boldsymbol{k}\eta,12}(t_1,t')+ \\
	&  G^{<}_{dd,22}(t,t_1) \Sigma^{a}_{d\beta\boldsymbol{k}\eta,22}  g^{a}_{\beta\boldsymbol{k}\eta,22}(t_1,t') + 
	G^{<}_{dd,21}(t,t_1) \Sigma^{a}_{d\beta\boldsymbol{k}\eta,11}  g^{a}_{\beta\boldsymbol{k}\eta,12}(t_1,t').
\end{split}
\end{equation}
Put above into Eq. \eqref{spincurr}, we get the spin-resolved currents
\begin{equation}\label{upcurr}
\begin{split}
	I_{\beta\uparrow}(t)= \frac{2e}{\hbar^2}\int dt_1 ~ \mathrm{Re} \sum_{\boldsymbol{k},\eta} |t_\beta|^2 [& G^{r}_{dd,11}(t,t_1)  g^{<}_{\beta\boldsymbol{k}\eta,11}(t_1,t) +  G^{<}_{dd,11}(t,t_1) g^{a}_{\beta\boldsymbol{k}\eta,11}(t_1,t) + \\
		& G^{r}_{dd,12}(t,t_1)  g^{<}_{\beta\boldsymbol{k}\eta,21}(t_1,t) +  G^{<}_{dd,12}(t,t_1) g^{a}_{\beta\boldsymbol{k}\eta,21}(t_1,t)]; \\
	I_{\beta\downarrow}(t)= \frac{2e}{\hbar^2}\int dt_1 ~ \mathrm{Re} \sum_{\boldsymbol{k},\eta} |t_\beta|^2 [& G^{r}_{dd,22}(t,t_1)  g^{<}_{\beta\boldsymbol{k}\eta,22}(t_1,t) +  G^{<}_{dd,22}(t,t_1) g^{a}_{\beta\boldsymbol{k}\eta,22}(t_1,t) + \\
		& G^{r}_{dd,21}(t,t_1)  g^{<}_{\beta\boldsymbol{k}\eta,12}(t_1,t) +  G^{<}_{dd,21}(t,t_1) g^{a}_{\beta\boldsymbol{k}\eta,12}(t_1,t)]
\end{split}
\end{equation}

For the advanced and lesser Green’s functions of the spin SC lead without coupling to the quantum dot, they can be solved under the wide-band approximation, which are given by
\begin{equation}
	\begin{split}
		&\sum_{\boldsymbol{k}} g^{a}_{\beta\boldsymbol{k}\eta}(t,t') \approx i\theta(t'-t) \rho_{N}(0)\int d\varepsilon ~
		\frac{e^{-i\varepsilon(t-t')/\hbar}}{\sqrt{\varepsilon^2-\Delta^2}}
		\left(
		\begin{array}{cc}
			|\varepsilon|&\Delta e^{i\phi_\beta} \frac{|\varepsilon|}{\varepsilon}\\
			\Delta e^{-i\phi_\beta} \frac{|\varepsilon|}{\varepsilon}&|\varepsilon|\\
		\end{array}
		\right); \\
		&\sum_{\boldsymbol{k}} g^{<}_{\beta\boldsymbol{k}\eta}(t,t') \approx i \rho_{N}(0)\int d\varepsilon ~ f(\varepsilon)
		\frac{|\varepsilon|}{\sqrt{\varepsilon^2-\Delta^2}} e^{-i\varepsilon(t-t')/\hbar}
		\left(
		\begin{array}{cc}
			1& \Delta e^{i\phi_\beta}\frac{1}{\varepsilon}\\
			\Delta e^{-i\phi_\beta}\frac{1}{\varepsilon}&1\\
		\end{array}
		\right),
	\end{split}
\end{equation}
where $\rho^{N}(0)$ is the density of states of a single lead in the norm state ($\Delta=0$) at the Fermi energy $E_f=0$.
Insert this equation into Eq. \eqref{upcurr}, take the Fourier transformation of the Green's functions from time domain to frequency domain and use the relation $G^{<}_{dd}(\varepsilon)=f(\varepsilon)[G^{a}_{dd}(\varepsilon)-G^{r}_{dd}(\varepsilon)]$, we obtain
\begin{equation}
\begin{split}
	I_{\beta\uparrow}(t)\approx& -\frac{4e}{\hbar^2} \mathrm{Im} \int \frac{d\varepsilon}{2\pi} ~ \Gamma \frac{|\varepsilon|}{2\sqrt{\varepsilon^2-\Delta^2}}
	f(\varepsilon) \left\{ [G^{a}_{dd,11}(\varepsilon)+G^{r}_{dd,11}(\varepsilon)]+\frac{\Delta e^{-i\phi_\beta}}{\varepsilon}[G^{a}_{dd,12}(\varepsilon)+G^{r}_{dd,12}(\varepsilon)] \right\}; \\
	I_{\beta\downarrow}(t)\approx& -\frac{4e}{\hbar^2} \mathrm{Im} \int \frac{d\varepsilon}{2\pi} ~ \Gamma \frac{|\varepsilon|}{2\sqrt{\varepsilon^2-\Delta^2}}
	f(\varepsilon) \left\{ [G^{a}_{dd,22}(\varepsilon)+G^{r}_{dd,22}(\varepsilon)]+\frac{\Delta e^{+i\phi_\beta}}{\varepsilon}[G^{a}_{dd,21}(\varepsilon)+G^{r}_{dd,21}(\varepsilon)] \right\},
\end{split}
\end{equation}
where $\Gamma=2\pi |t_\beta|^2 \rho^N(0)$ is the linewidth function of the lead. When $\Delta=0$, the first bracket in the above equations represents the current tunneling from normal metal to norm metal at equilibrium, which should be zero as indeed. For $\Delta\neq 0$, this term contributes small and can be dropped.
Therefore, the spin-resolved currents reduce to
\begin{equation}\label{updowncurr}
	\begin{split}
		&I_{\beta\uparrow}(t)\approx -\frac{4e}{\hbar^2} \mathrm{Im} \int \frac{d\varepsilon}{2\pi} ~ \Gamma^{\beta} \frac{\Delta e^{-i\phi_\beta}}{2\sqrt{\varepsilon^2-\Delta^2}} \frac{|\varepsilon|}{\varepsilon}
		f(\varepsilon) [G^{a}_{dd,12}(\varepsilon)+G^{r}_{dd,12}(\varepsilon)]; \\
		&I_{\beta\downarrow}(t)\approx -\frac{4e}{\hbar^2} \mathrm{Im} \int \frac{d\varepsilon}{2\pi} ~ \Gamma^{\beta} \frac{\Delta e^{+i\phi_\beta}}{2\sqrt{\varepsilon^2-\Delta^2}} \frac{|\varepsilon|}{\varepsilon}
		f(\varepsilon) [G^{a}_{dd,21}(\varepsilon)+G^{r}_{dd,21}(\varepsilon)],
	\end{split}
\end{equation}

To evaluate the retarded Green's function $G^{r}_{dd}(\varepsilon)$ of the quantum dot, let us first calculate the Green's function $g^{r}_{dd}(\varepsilon)$ of the dot without coupling to the leads. It can be given by
\begin{equation}
	g^{r}_{dd}(\varepsilon)=\left[(\varepsilon+i0^{\dagger})I-\left(
	\begin{array}{cc}
		\varepsilon_{0\uparrow}& 0\\
		0&\varepsilon_{0\downarrow}\\
	\end{array}
	\right)\right]^{-1}=\left(
	\begin{array}{cc}
		\frac{1}{\varepsilon-\varepsilon_{0\uparrow}+i0^{\dagger}}& 0\\
		0&\frac{1}{\varepsilon-\varepsilon_{0\downarrow}+i0^{\dagger}}\\
	\end{array}
	\right).
\end{equation}
And the retarded self-energies of the quantum dot contributed by the leads are
\begin{equation}
	\Sigma^{r}(\varepsilon)=\sum_{\beta} \Sigma^{r}_{\beta}(\varepsilon) = \sum_{\beta} |t_\beta|^2 \sum_{\boldsymbol{k},a} g^{r}_{\beta\boldsymbol{k}a\beta\boldsymbol{k}a}(\varepsilon) = -i \sum_{\beta} \frac{\Gamma}{\sqrt{\varepsilon^2-\Delta^2}}
	\left(
	\begin{array}{cc}
		|\varepsilon|&\Delta e^{i\phi_\beta} \frac{|\varepsilon|}{\varepsilon}\\
		\Delta e^{-i\phi_\beta} \frac{|\varepsilon|}{\varepsilon}&|\varepsilon|\\
	\end{array}
	\right).
\end{equation}
Then, the Green's function of the quantum dot with coupling to the leads can be given by
\begin{equation}
	G^{r}_{dd}(\varepsilon)=\left[ (\varepsilon+i0^{\dagger})I-\left(
	\begin{array}{cc}
		\varepsilon_{0\uparrow}& 0\\
		0&\varepsilon_{0\downarrow}\\
	\end{array}
	\right) - \Sigma^{r}(\varepsilon) \right]^{-1}= \frac{1}{B(\varepsilon)}
	\left(
	\begin{array}{cc}
		g^{r,-1}_{00,22}-\Sigma^{r}_{22}& \Sigma^r_{12}\\
		\Sigma^r_{21}&g^{r,-1}_{00,11}-\Sigma^{r}_{11}\\
	\end{array}
	\right),
\end{equation}
where $B(\varepsilon)=(	g^{r,-1}_{00,11}-\Sigma^{r}_{11})(	g^{r,-1}_{00,22}-\Sigma^{r}_{22})-\Sigma^r_{12}\Sigma^r_{21}$. Bring $G^{r}_{dd}(\varepsilon)$ into Eq. \eqref{updowncurr} and use $G^{a}_{dd}(\varepsilon)=[G^{r}_{dd}(\varepsilon)]^{\dagger}$, we arrive at the final form of the equilibrium spin-resolved currents, flowing in lead $\beta=L$
\begin{equation}
	\begin{split}
		&I_{L\uparrow}\approx  \frac{4e}{\hbar} \int \frac{d\varepsilon}{2\pi} ~ \frac{\Gamma^2\Delta^2}{\varepsilon^2-\Delta^2}f(\varepsilon) \mathrm{Im} \frac{1}{B^{\ast}(\varepsilon) }\mathrm{sin}(\phi_R-\phi_L); \\
		&I_{L\downarrow}\approx  \frac{4e}{\hbar} \int \frac{d\varepsilon}{2\pi} ~ \frac{\Gamma^2\Delta^2}{\varepsilon^2-\Delta^2}f(\varepsilon) \mathrm{Im} \frac{1}{B^{\ast}(\varepsilon) }\mathrm{sin}(\phi_L-\phi_R),
	\end{split}
\end{equation}
which is just Eq. \eqref{Lspincurr} presented in the main text.

\end{appendices}

\twocolumngrid
\bibliographystyle{apsrev4-1}
\bibliography{refspinSC}

\end{document}